\pdfoutput=1
\documentclass[a4paper,fleqn,usenatbib]{mnras}
\usepackage{graphicx}	% Including figure files
\usepackage{amsmath}	% Advanced maths commands
\usepackage{amssymb}	% Extra maths symbols

\usepackage{txfonts}

\usepackage[T1]{fontenc}
\usepackage{ae,aecompl}

\title[Coevolution of Binaries and Gaseous Discs]{Coevolution of Binaries and Circumbinary Gaseous Discs}

\author[Fleming \& Quinn]{
David P. Fleming\thanks{E-mail: dflemin3@uw.edu (DPF)}
and Thomas R. Quinn
\\
Department of Astronomy, University of Washington, Seattle, WA 98105, USA\\
}

\date{Accepted XXX. Received YYY; in original form ZZZ}

\pubyear{2016}

\begin{document}
\label{firstpage}
\pagerange{\pageref{firstpage}--\pageref{lastpage}}
\maketitle

\begin{abstract}
The recent discoveries of circumbinary planets by $\it Kepler$ raise questions for contemporary planet formation
models.  Understanding how these planets form requires characterizing their formation environment, the circumbinary protoplanetary disc,
and how the disc and binary interact and change as a result.  The central binary excites resonances in the surrounding
protoplanetary disc that drive evolution in both the binary orbital elements and in the
disc.  To probe how these interactions impact binary eccentricity and disk structure evolution, N-body smooth particle
hydrodynamics (SPH) simulations of gaseous protoplanetary discs
surrounding binaries based on Kepler 38 were run for $10^4$ binary
periods for several initial binary eccentricities.  We find that
nearly circular binaries weakly couple to the disc via a parametric
instability and excite disc eccentricity growth.  Eccentric binaries strongly couple to the disc
causing eccentricity growth for both the disc and binary.
Discs around sufficiently eccentric binaries that strongly couple to the disc develop an
$m = 1$ spiral wave launched from the 1:3 eccentric outer Lindblad
resonance (EOLR) that corresponds to an alignment of gas
particle longitude of periastrons. All systems display binary semimajor axis decay due to dissipation from the
viscous disc.  \end{abstract}

\begin{keywords}
binaries: general -- hydrodynamics -- protoplanetary discs
\end{keywords}

%%%%%%%%%%%%%%%%%%%%%%%%%%%%%%%%%%%%%%%%%%%%%%%%%%

%%%%%%%%%%%%%%%%% BODY OF PAPER %%%%%%%%%%%%%%%%%%

\section{Introduction}

Observations by $Kepler$ of binary systems have so far yielded 11 transiting circumbinary planets.  Until recently, all discovered circumbinary planets have resided near to or outside the dynamical stability limit characterized by \citet{Dvorak86,HolmanWiegert99}.  This
finding has prompted many to study how planets form in such systems and why they seem biased towards lying at the brink of dynamical instability \citep{Welsh14,PierensNelson13}.  Simulations and theoretical arguments by \citet{Pelupessy13} and \citet{Bromley15} rule out in situ formation
in the inner edge of the disc near where several circumbinary planets have been observed.  Currently, one of the most successful models in explaining the circumbinary planet population is planetary migration.  
In this model, a planet, having formed farther out in the disc, migrates inward through viscous interactions with a gaseous disc, potentially undergoing planet-planet scattering, until it reaches near its observed location.  Several previous studies applied migration to observed circumbinary planetary systems and have been able to show that for certain disc and viscous drag models, the observed planet orbits are nearly recovered \citep[e.g.][]{Kley14,KleyHag15,PierensNelson07,PierensNelson13}.

However, the recent discovery by \citet{Kostov15} of the first long-period transiting circumbinary planet complicates this picture.  The newly discovered planet candidate, KOI-2939b, is a Jupiter-sized exoplanet on a roughly 3 year orbit that suggestively lies in the conservative habitable zone of two short-period G dwarfs \citep{Kostov15}.  The existence of a such a system shows that not all circumbinary planets are driven, either through migration or other mechanisms, inward towards the dynamical stability limit.  This has important consequences for not only how planets migrate and evolve in a viscous circumbinary disc, but also for how and where these planets form within the protoplanetary disc.

Clearly, characterizing the influence exerted on the planet by the protoplanetary disc is important for understanding how such systems form.  Additionally, the influence of the central binary on the external disc causes the disc to evolve and potentially undergo large scale changes.  Previous studies of both protoplanetary discs and accretion discs around binary supermassive black holes (SMBHs) found that circumbinary discs can become eccentric, precess, and have density waves excited from resonances \citep{Dunhill15,MacFadyen08,Papaloizou01,PierensNelson07,PierensNelson13,Roedig12}.  Specifically in the context of planetary systems, a hydrodynamic theory presented by \citet{Lubow91} showed that nonlinear coupling mediated by density waves launched at eccentric Lindblad resonances causes disc eccentricity growth. 

Much theoretical work in this area has focused on how planetesimals
can grow and evolve in circumbinary discs.  \citet{Paardekooper12}
found that for planetesimals on circumbinary orbits, in situ formation
proves quite difficult, suggesting that planets form far out in the
disc and subsequently migrate inward.  More recently, \citet{Bromley15}
showed that planetesimals on the most circular orbit about the central
binary can attain small relative velocities, facilitating their growth.
With a model that considers both the gravity and gas drag of a
precessing, eccentric circumbinary disc, \citet{Silsbee15a} found that
the inner radius for $10-100$m planetesimal growth depends on the disc eccentricity, density and precession rate.  The binary's influence on the disc is not without a cost, however, as the disc also drives changes in the binary orbital elements, which as they evolve, can change how the binary influences the disc.

Early work on how binary systems interact with accretion discs by \citet{GT80} studied Jupiter's interaction with the Sun's protoplanetary disc.  This study showed that a satellite's orbital eccentricity could be increased through energy and momentum transfers with the disc at Lindblad resonances, causing significant changes over a few thousand years.  The case of accretion discs around two objects was explored by \citet{Pringle91} and later by \citet{Papaloizou01}.  \citet{Pringle91} found that a central binary's interaction with an external accretion disc can decrease binary separation and change binary eccentricity, depending on disc structure.  In the context of a Jupiter to brown dwarf mass companion orbiting a central star,  \citet{Papaloizou01} found that for sufficiently massive companions, a coupling between small initial disc eccentricity and the companion's
tidal potential excited an $m = 2$ wave from the 1:3 EOLR causing further disc eccentricity growth.  Wave excitation at the 1:3 EOLR can also couple with the interaction between the eccentric disc and the companion to induce orbital eccentricity growth of the companion.  

Many recent and past works explored the more general case of binary stars evolving under the influence of a gaseous circumbinary disc.  Simulations of unequal mass binaries embedded in a protoplanetary disc by \citet{Arty91} show rapid binary eccentricity growth and semimajor axis decay due to interactions with the 1:3 EOLR and the viscous disc.  Subsequent theoretical work by \citet{Arty96b} and \citet{Arty2000} derived relations to show how resonant and viscous interactions drive $\dot{e}_{bin}$ and $\dot{a}_{bin}$.  Binary orbital element and disc evolution has also been explored in systems with circumbinary gaseous discs containing migrating planets in simulations by \citet{PierensNelson07} who also found binary and disc eccentricity growth.

Characterizing disc-binary interactions are important on much larger scales, as well.  Simulations by \citet{Mayer07} show that eccentric supermassive
black hole binaries can rapidly form from the merger between two spiral galaxies.  An external disc forms exterior to the SMBH binary
as interactions with the disc and external gas clouds can cause the black hole separation to decrease \citep{Roskar15}.  Numerous studies have been 
conducted to explore how binary SMBH-disc interactions cause binary SMBH eccentricity growth and semimajor axis decay, possibly 
explaining the ``last parsec" problem allowing SMBH binaries to coalesce \citep[e.g.][]{Armitage05,Escala05,Cuadra09,Roedig12,Aly15}.  Extensive work
has also been made to explore accretion onto binary SMBHs, how it varies with both the binary and disc properties and what impact accretion
has on the dynamical properties of the system \citep[e.g.][]{Shi12,DOrazio13,Nixon13,Farris14}.
 
For circumbinary systems, the binary and its disc are intertwined in
non-trivial ways.  The coevolution of such systems depends strongly on
resonant interactions that can impart significant changes on both the
binary and disc.  Previous simulations focused on studying either disc evolution or binary evolution, often making approximations
such as holding the binary orbital elements fixed.  In this work we seek to explore how the coevolution
of the disc and binary proceeds by allowing all particles to gravitationally interact.  
Using the Kepler 38 binary as our model system, we present the results of N-body
SPH simulations of unequal mass binaries of variable initial eccentricity 
that probe the dynamical coupling of binary stars and a circumbinary disc.  
We show that the initial eccentricity of the binary dictates how strongly the disc-binary system resonantly couples.
The strength of this coupling in turn dictates how eccentricity grows in either the disc, the binary or both and how structure 
forms within the disc.

\section{Simulations}

\subsection{Methods} \label{methods_section}

The simulations described in this paper were performed in 3D using the massively parallel N-body 
and Smooth Particle Hydrodynamics (SPH) code, ChaNGa
\citep{Menon15}.\footnote{A public version of ChaNGa is available
  from {\tt http://www-hpcc.astro.washington.edu/tools/changa.html}.}
ChaNGa, implemented in Charm++ \citep{KaleKrishnan96}, uses a modified version of the Barnes-Hut tree algorithm with hexadecapole 
order multiples and a node opening criterion of $\theta_{BH} =
  0.7$ for fast and accurate calculation of gravitational forces.  The Euler equations which describe
the gas dynamics of the simulated circumbinary disc were solved using an SPH implementation based on
\citet{Wadsley04}.  ChaNGa uses a multistepping algorithm that gives
each particle its own timestep to ensure sufficient dynamical
resolution \citep{Quinn97}.  Artificial viscosity was implemented using
Monaghan viscosity \citep{Monaghan83}.  The viscosity $\alpha_{SPH}$
  and $\beta_{SPH}$ parameters were set to 1 and 2, respectively.  The Balsara switch was used to limit shear viscosity \citep{Balsara95}.  For additional information about the implementation and performance of ChaNGa see \citet{Jetley08}.

% Table 1: Kepler 38 parameters
\begin{table}
	\centering
	\caption{Kepler 38 parameters adapted from \citet{Orosz12}.}
	\begin{tabular}{cccc} % four columns, alignment for each
		\hline
		m$_1$ [M$_{\odot}$] & m$_2$ [M$_{\odot}$] & $a_{bin}$ [AU] & $e_{bin}$\\
		\hline
		0.949 & 0.249 & 0.1469 & 0.1032\\
		\hline
	\end{tabular}
	\label{tab:table_1}
\end{table}

The orbits of both the binary and all the disc particles are computed using ChaNGa's implementation of the 
symplectic leapfrog integrator.  Each particle feels the force of gravity due to every other particle, including
the two stars.  Since the orbits of the stars are not integrated using a higher order technique, we employed conservative
time stepping.  We ran simulations shrinking time stepping parameters to 
ensure that the binary orbit was accurately resolved such that any evolution in the orbital parameters is due to 
gravitational interactions with the disc and not any numerical
effects.  The timestep picking criterion used was {\tt bGravStep} where  
$\Delta t = \eta/\sqrt{r/a}$ where $a$ is the acceleration of the
particle, $r$ is the distance to the particle or node that causes the largest
acceleration and $\eta$ is an accuracy parameter.  We used this timestep criterion as its form is particularly suited for Keplerian orbits, and in the absence of collisions, it has the desirable property of giving the particle a fixed number of timesteps per orbital period ($n = 2\pi/ \eta$) \citep{Richardson2000}.  We found that 
$\eta = 0.005$ yielded sufficiently small timesteps to properly
resolve the dynamics of the binary.  In addition, we used a Courant number of
0.4 for enforcing the Courant condition.

All simulations were ran on the University of Washington's Hyak supercomputer cluster.  Each simulation was ran on either a 12 or 16 core node for about a month each with roughly 8,000 core hours used per simulation on average.  In total, over 100,000 core hours were used for the entire suite of simulations.

\subsection{Model Parameters}

The model system studied in this paper is Kepler 38, a binary composed of a G and an M dwarf.  The physical parameters for this 
system were adapted from \citet{Orosz12} and are given in Table~\ref{tab:table_1}.  We ran eight simulations of Kepler 38 embedded in a 
circumbinary disc for $10^4$ binary orbits, about 520 years.  The simulations considered in this paper were ran primarily to examine how 
the surrounding circumbinary disc impacts the dynamics of the central
binary as a function of the eccentricity of the binary, $e_{bin}$, and
how the disc in turn evolves.  The disc masses are also varied in Simulations 6, 7, and 8 to explore its role in the evolution of the binary's orbital parameters.  
We study the sensitivity of our results on disk resolution and aspect ratio in Simulations 9, 10, and 11 with discussion in Section 
\ref{VaryingDiscProps}.  A summary of the relevant simulation parameters used in this study are given in Table~\ref{tab:table_2}. 

\subsection{Initial Conditions}

Initial conditions for the circumbinary disc were computed using the
Python package {\em diskpy}\footnote{{\em diskpy} GitHub repository: https://github.com/ibackus/diskpy}.  
Given the stellar and disc parameters, {\em diskpy}
calculates the positions and velocities for gas particles for a
protoplanetary disc in equilibrium.   Each gas particle is placed on a
circular orbit about the the binary's barycenter assuming a central
mass equal to the total binary mass.  
The gas particles' orbits also feel the force of radial pressure gradients within the disc.  Great care was taken to ensure that the disc
was as close to equilibrium as possible in both the radial and vertical directions to avoid any influence from an initial disequilibrium
state.  For a more in-depth description of the initial conditions generated by diskpy, see \citet{diskpy}.

The disc, composed of $10^5$ SPH particles unless stated otherwise, has initial inner and outer boundaries of 0.25 and 4.0 AU,
respectively.  The initial surface density profile of the disc is set to 
%% Sigma(r)
\begin{equation}
\Sigma(r) = \Sigma_0 r^{-1/2}
\end{equation}
 where $\Sigma_0$ is assigned such that if the disc extended to 10 AU, it would have a total mass of about 0.01 
M$_{\odot}$, similar to the models of \citet{PierensNelson07} and \citet{Kley14}.  To prevent numerical 
artefacts, the disc inner and outer edges of the surface density profiles are smoothed.  For the inner edge, a smooth polynomial spline approximation to a step function is used while the outer edge of the disc is given by an exponential cutoff.
 
An open boundary condition is applied at the inner disc edge such that inflowing particles are allowed to be accreted
by the central binary.  The stars, modeled as ``sink'' particles, accrete a gas particle if it enters the Roche lobe of a star.  This process is modeled by adding the mass of the accreted gas particles to the star's mass and conserving linear momentum throughout the accretion process.  Unless stated otherwise, the 
sink radius was set to 0.066 AU.  A simulation was ran with a sink radius a factor of 5 smaller than the typical value and we found that our results were not 
affected.
 
The disc has a locally isothermal temperature profile following 
%% T(r)
\begin{equation}
\label{eqn:disc_temp_profile}
T(r) = T_0(r_0/r)
\end{equation}
where at 1 AU from the barycenter, $T_0$ is
750 K.  Unless otherwise specified, the circumbinary discs are initialized to be stable against axisymmetric perturbations with Q$_{min} > 1.5$ \citep{Toomre64}.
The vertical scale height of the disk, $H$, was on average resolved by
  2.5 resolution lengths, $h$, where $2 h$ is the distance
to the nearest 32 neighboring SPH particles.

%%%% Figure 1: disc surface density vs binary eccentricity
\begin{figure}
	\includegraphics[width=\columnwidth]{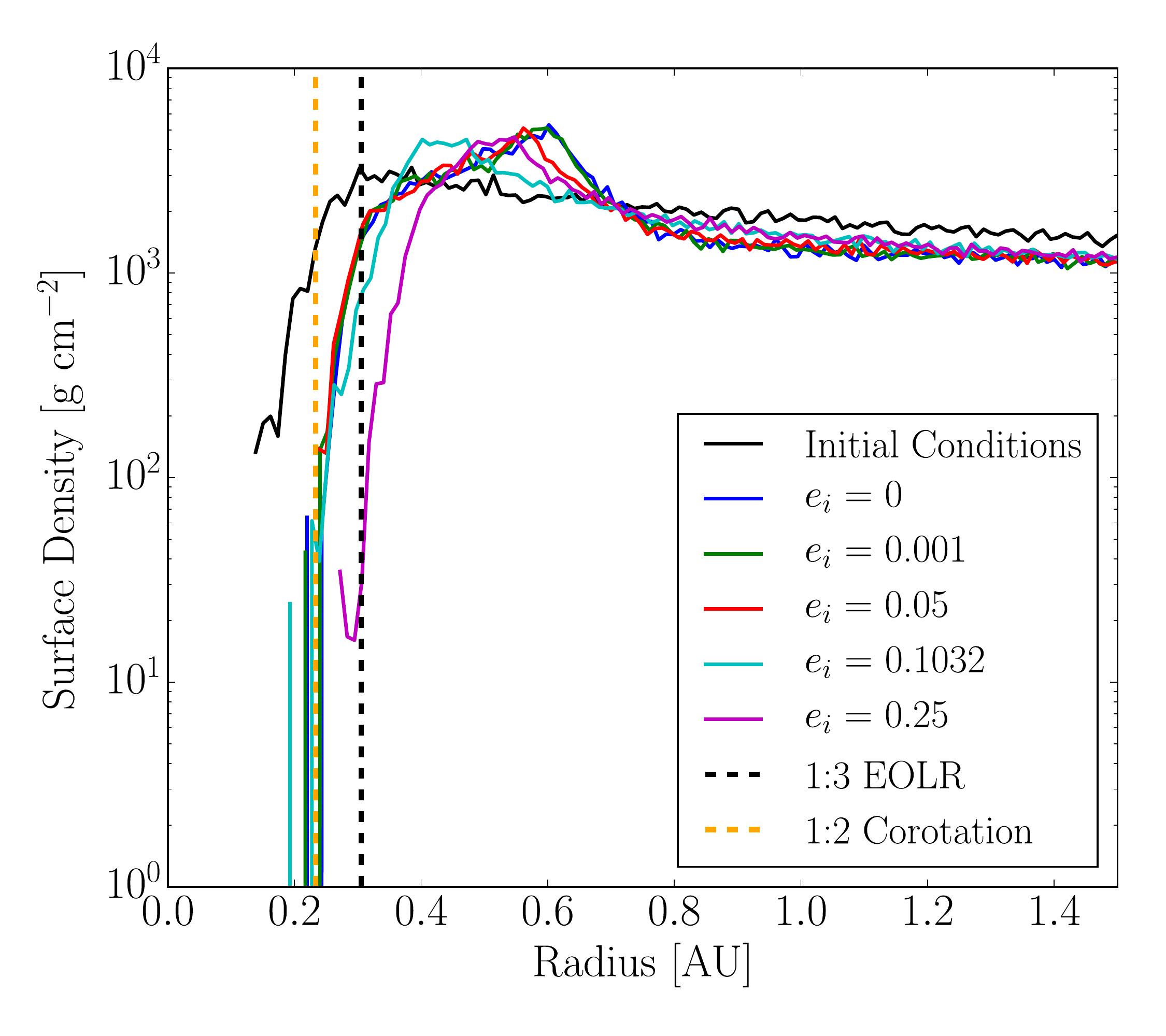}
    \caption{Azimuthally averaged surface density profiles for Simulations 1-5 after 200 years of evolution.  Overplotted are the 1:3 EOLR (dashed black line) and the 1:2 corotation resonance (dashed orange line).}
    \label{fig:figure1}
\end{figure}

% Table 2: Simulations
\begin{table}
	\centering
	\caption{Simulation parameters.}
	\begin{tabular}{ccccc} % five columns, alignment for each
		\hline
		Simulation Number & $e_{bin}$ & $M_{disc}$ [M$_{\odot}$] & N$_{gas}$ & H/R \\
		\hline
		1 & 0 & 0.00383 & $10^5$ & 0.06 \\
		2 & 0.001 & 0.00383 & $10^5$ & 0.06 \\
		3 & 0.05 & 0.00383 & $10^5$ & 0.06 \\
		4 & 0.1032 & 0.00383 & $10^5$ & 0.06 \\
		5 & 0.25 & 0.00383 & $10^5$ & 0.06 \\ 
		6 & 0.1032 & 0.00766 & $10^5$ & 0.06 \\
		7 & 0.1032 & 0.00192 & $10^5$ & 0.06 \\
		8 & 0.1032 & 0.00574 & $10^5$ & 0.06 \\
		9 & 0.1032 & 0.00383 & 5 $\times$ $10^4$ & 0.06 \\
		10 & 0.1032 & 0.00383 & 2 $\times$ $10^5$ & 0.06 \\
		11 & 0.1032 & 0.00383 & $10^5$ & 0.12 \\
		\hline
	\end{tabular}
	\label{tab:table_2}
\end{table}

\section{Results and Analysis}

	%% Gap Clearing Subsection %%

\subsection{Gap Clearing} \label{GapClearing}

As the simulation evolves, the time varying gravitational force of the
binary truncates the inner edge of the circumbinary disc and excites
various Lindblad resonances within the disc.  The gap, cleared quickly
in about 100 binary orbits, is preserved by a balance of resonant and
viscous torques within the disc \citep{Arty94}.  For larger gaps,
corotation and Lindblad resonances can fall within the evacuated
region removing their influence from the system.  These resonances,
especially those closest to the inner edge of the disc, can drive
evolution in the binary's orbital elements \citep{GT80,Arty91} and
pump eccentricity in the disc \citep{Papaloizou01}.  Therefore, in
order to understand the subsequent dynamical evolution of both the
binary and the disc, the inner disc edge structure must be understood to see which
resonances may play a role.  The approximate size of the gaps found in
these simulations is $r \approx 2 a_{bin}$ with more eccentric binaries
producing larger gaps, in good agreement with the results of
\citet{Arty94}.  Our results are also consistent with the findings of \citet{DOrazio16} who show that for binary mass ratios above $q = 0.04$ as is the case 
for our simulations, a hollow central cavity forms around the secondary within the circumbinary disc.  Gaps of this size tend to remove the eccentricity-damping 
1:2 corotation resonance while leaving the 1:3 EOLR as the closest to the inner disc edge at a radius of $2.08 a_{bin}$ suggesting that this resonance 
dominates the evolution of the binary's eccentricity and semimajor axis \citep{Arty91,Arty92,Papaloizou01}.

To verify the absence of the 1:2 corotation resonance and prominence
of the 1:3 EOLR, the surface density profiles for Simulations $1-5$
are shown in Fig.~\ref{fig:figure1} after 200 years of evolution.
For all $e_{bin}$, the surface density at the 1:3 EOLR is at least an
order of magnitude larger than at the 1:2 corotation resonance when it
is present within the disc.  Since the gap opens rather quickly, in of order 5 years for the Kepler 38 binary,  this finding supports the hypothesis that the 1:3 EOLR will be the dominant resonance within the disc that will drive subsequent evolution within the system.

	%% disc Evolution Subsection %%

\subsection{Disc Evolution}

The structure and eccentricity of the protoplanetary disc was examined over $10^4$ binary periods in over 230 snapshots
for each simulation to track how it evolves with the central binary.

%%% Figure 2: disc eccentricity vs time
\begin{figure}
	\includegraphics[width=\columnwidth]{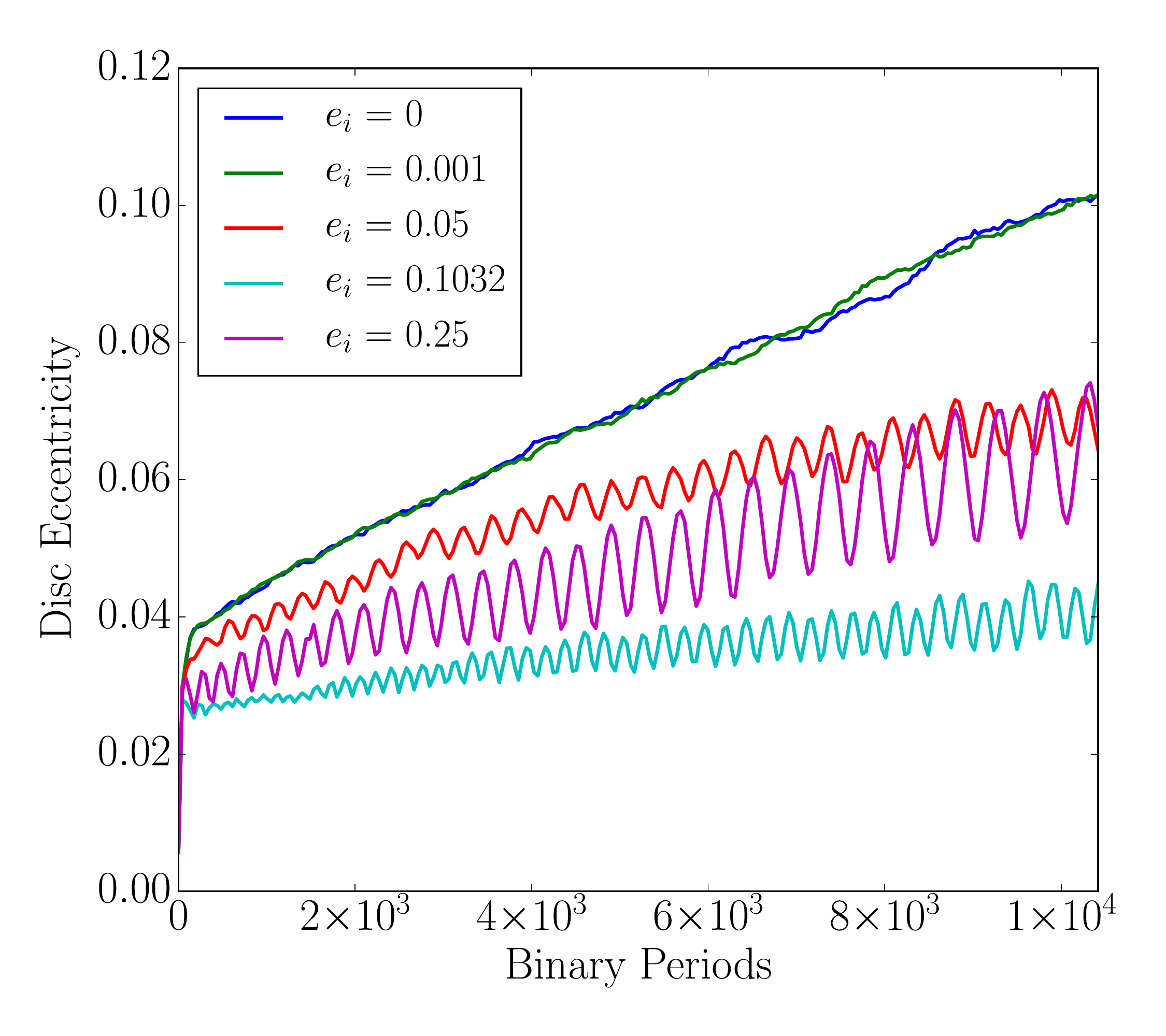}
    \caption{Disc eccentricity versus simulation time in units of binary orbits as a function of initial binary eccentricity.}
    \label{fig:figure2}
\end{figure}

		%% disc Eccentricity Evolution Subsubsection %%

\subsubsection{Disc Eccentricity Growth} \label{discEccEvolution}

A central binary excites eccentricity in the surrounding
circumbinary disc via resonant gravitational interactions
\citep{Papaloizou01,Arty96a}.  Previous simulations of circumbinary
discs \citep{Kley08,Papaloizou01,PierensNelson07,PierensNelson13} found
that disc eccentricity increases due to interactions that occur at
the 1:3 EOLR.  To explain why the disc becomes eccentric, \citet{Papaloizou01} showed that disc eccentricity growth occurs via a parametric instability driven by coupling between the binary's tidal potential and a disc $m = 1$ mode due to a small initial disc eccentricity.  This coupling excites an $m = 2$ spiral wave from the 1:3 EOLR that removes angular momentum from the disc at constant energy making the gas orbits eccentric.  The material at the 1:3 EOLR rotates more slowly that the orbital pattern speed allowing the resonant torques to grow eccentricity in the system through the $m = 2$ wave.  We therefore expect our circumbinary discs to become eccentric as well.  To explore this effect in our simulations, we computed the disc eccentricity via a mass average following the prescription of \citet{PierensNelson07}

%% disc eccentricity equation from Pierens + Nelson 2007
\begin{equation}
\bar{e}_d = \frac{\int_0^{2 \pi} \int^{R_{out}}_{R_{in}} e \Sigma r dr d\phi }{\int_0^{2 \pi} \int^{R_{out}}_{R_{in}} \Sigma r dr d\phi},
\end{equation}
where $\Sigma$ is the local surface density and the integral was
evaluated out to a radius of 3 AU over 50 radial bins.  The disc eccentricity in each radial bin was taken to be the mass-weighted average of the eccentricity of all gas particles within the bin assuming the particles orbit the system's barycenter.  We neglected the influence of gas pressure in this calculation.

For all simulations, disc eccentricity growth is observed and consistent with the results
of the similar gaseous circumbinary disc simulations of \citet{Kley08}, \citet{PierensNelson07,PierensNelson13} and \citet{Farris14}.  The disc eccentricity 
change over time for each simulation is shown in Fig.~\ref{fig:figure2}. For $e_{bin} \approx 0$, significant disc eccentricity growth occurs.  
After about 500 years, the disc reaches eccentricities of about 0.1 while continuing to grow linearly.  For larger initial $e_{bin}$ up 
to $e_{bin} \approx 0.1$, less eccentricity growth occurs indicating that more eccentric binaries tend to produce less eccentric discs.

For non-zero $e_{bin}$, the disc eccentricity grows linearly with a
periodic modulation.  The period of this disc eccentricity oscillation
is similar to the inner disc edge clump precession timescale discussed
in Section \ref{discStructEvolution} below suggesting that the clump impacts the disc's eccentricity modulation, but only when the binary is sufficiently eccentric.  This finding is consistent with the work of \citet{Arty2000} who explain that the inner disc edge precesses with a period of $10^2 - 10^3$ binary periods when $e_{disc}/e_{bin} \sim 0.2-0.7$.  

\begin{figure}
	\includegraphics[width=\columnwidth]{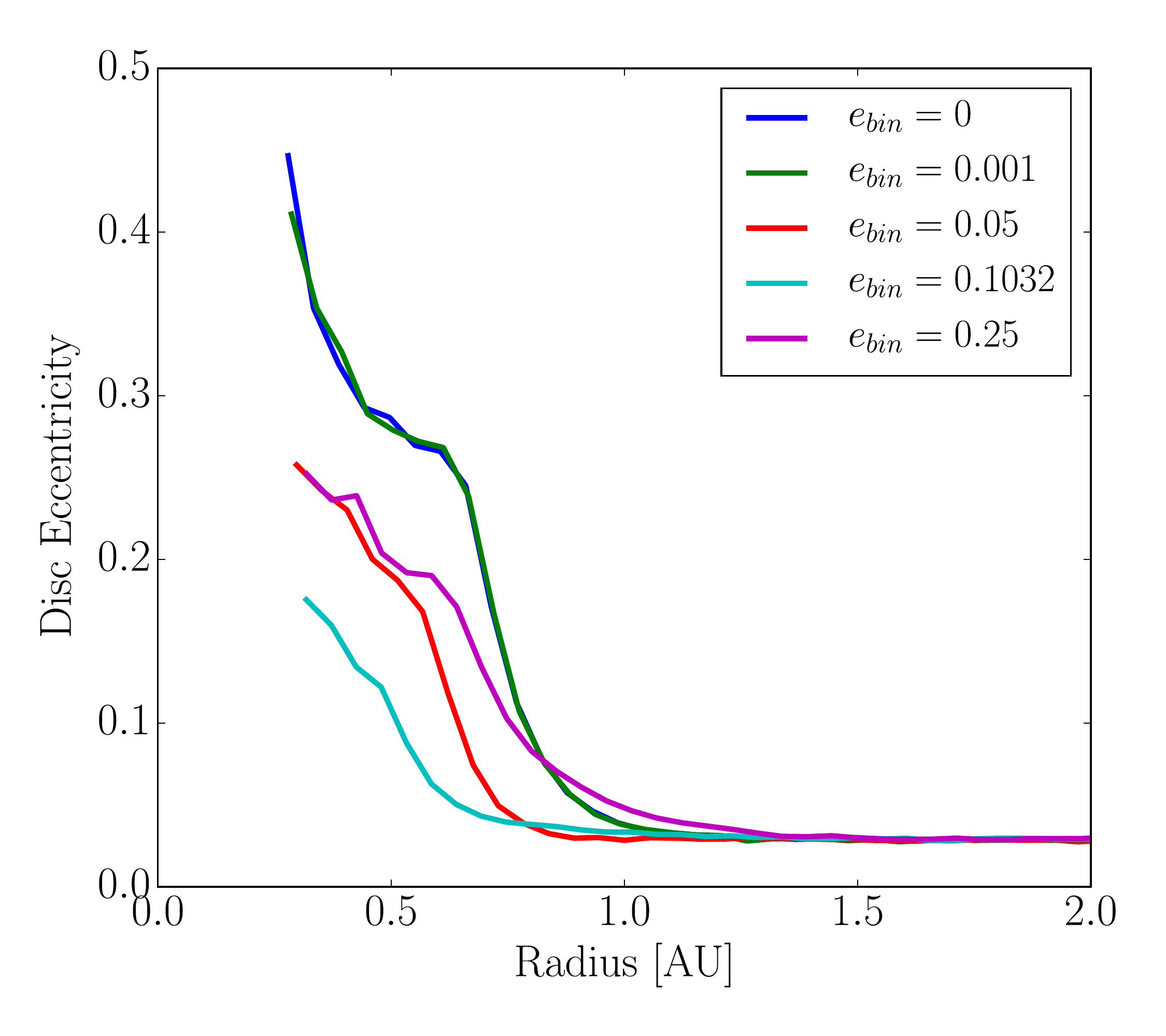}
    \caption{Disc eccentricity versus radius after about 520 years of evolution.}
    \label{fig:figure3}
\end{figure}

We find that the global disc eccentricity growth is predominantly due to the inner edge of the disc becoming eccentric.  Fig.~\ref{fig:figure3} shows the disc eccentricity versus radius for Simulations 1-5 after about 520 years of evolution.  For binaries with $e_{bin} \approx 0$, disc inner edge eccentricities are of order 0.4 while more eccentric binaries tend to produce less eccentric inner edges.

%%% Figure 3: disc eccentricity vs radius

To understand the eccentricity growth of our circumbinary discs,
specifically why less eccentric binaries tend to produce more
eccentric discs, we turn to the theory of \citet{Papaloizou01}.
\citet{Papaloizou01} explains that a nonlinear coupling between the
binary and a small initial disc eccentricity excites an $m = 2$ wave
from the 1:3 EOLR within the disc with a resonant forcing pattern
speed $\omega/2$ for binary orbital frequency $\omega$.  This wave
transports angular momentum outwards, driving eccentricity growth in
the system.  We confirmed the presence of the $m = 2$ spiral wave
originating from the 1:3 EOLR in our simulations via a Fourier transform over azimuthal
angle of the disc surface density.  This wave removes angular momentum from the disc, accounting for the increase in disc eccentricity.  

\begin{figure}
	\includegraphics[width=\columnwidth]{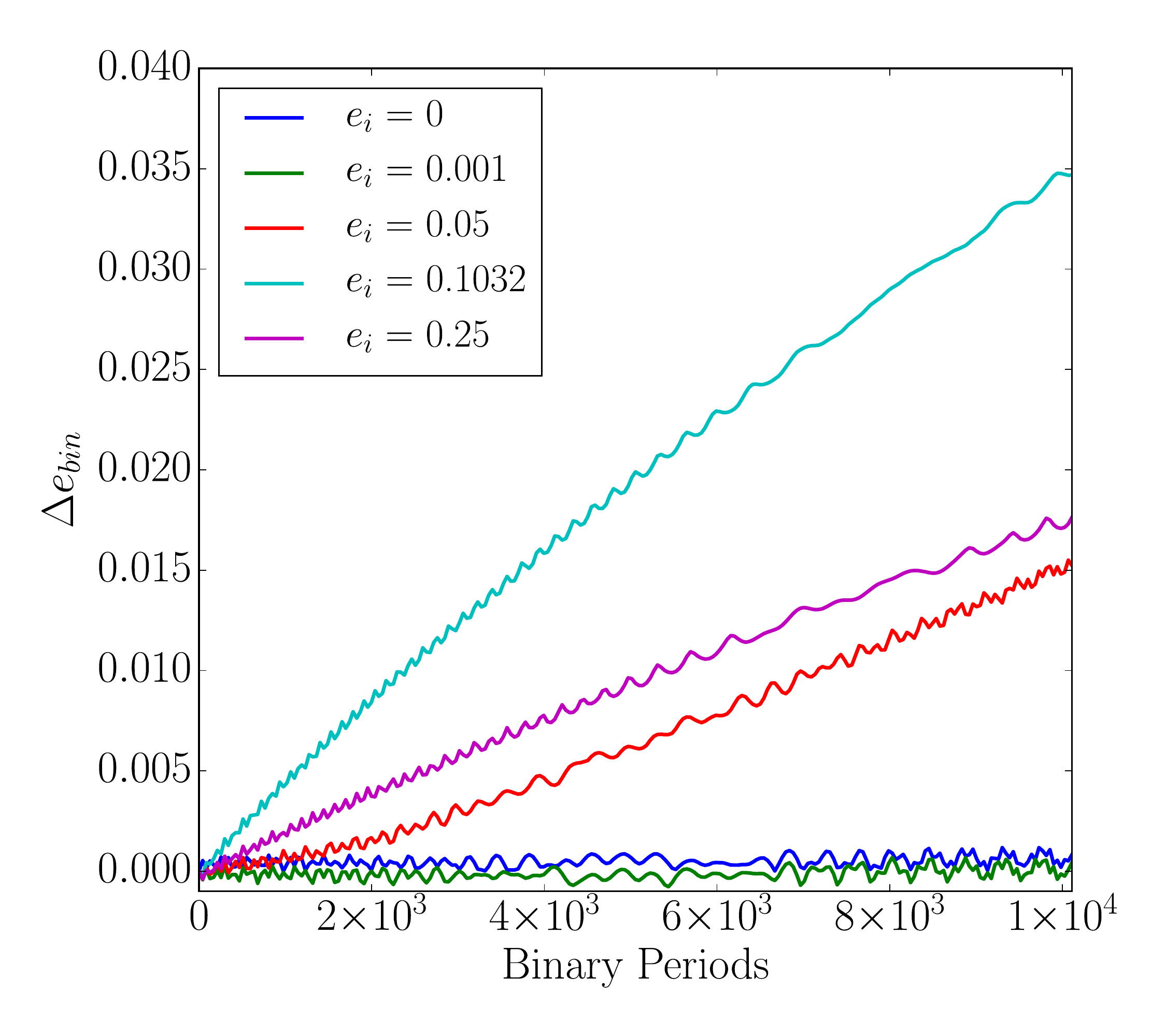}
    \caption{Change in binary eccentricity versus simulation time in units of binary orbits for several initial
binary eccentricities. For an initially circular and nearly circular binary, effectively no eccentricity growth occurs whereas 
for higher initial eccentricities, significant growth occurs.}
    \label{fig:figure4}
\end{figure}

Disc eccentricity growth as a function of $e_{bin}$ depends on how strongly the binary couples to the disc.  When strong coupling occurs, both the binary and inner disc edge grow to similar eccentricities while weak coupling results in the less massive of the binary or the disc inner edge developing appreciable eccentricity.  For initially circular binaries, \citet{Papaloizou01} described strong coupling as occurring when the disc mass within the gap radius is comparable to the mass of the secondary.  For our simulations, the mass of the secondary is roughly an order of magnitude larger than the mass of the entire disc, so we expect the binary to be weakly coupled to the disc resulting in significant disc eccentricity growth as is observed in our simulations (see Fig.~\ref{fig:figure2} and Fig.~\ref{fig:figure3}).  For initially eccentric binaries, the \citet{Papaloizou01} strong coupling criterion does not apply.  Instead, we note that the time-averaged orbit of an eccentric binary corresponds to an azimuthal $m = 1$ mode perturbation to a circular orbit.  This $m = 1$ mode couples to the $m = 1$ mode of the eccentric inner disc edge, placing the system in the strong coupling regime causing the disc eccentricity and $e_{bin}$ to grow to similar magnitudes, as is observed in our simulations (see Fig.~\ref{fig:figure2} and Fig.~\ref{fig:figure4}).  We discuss these coupling mechanisms and their implications in more detail in Sections \ref{discStructEvolution} and \ref{BinaryEvolution}.

%%% Figure 4:Binary eccentricity vs time

The $e_{bin} = 0.25$ and $e_{bin} = 0.05$ simulations show larger disc eccentricity than
the $e_{bin} = 0.1032$ case in contrast to expected behavior.  For
$e_{bin} = 0.05$, this can be understood as intermediate coupling.
The initial binary eccentricity is not low enough to conform exactly
to the \citet{Papaloizou01} criterion and is not large enough to
couple strongly to the disc inner edge resulting in an intermediate
coupling with larger disc eccentricity growth than the $e_{bin} = 0.1032$ case and also
appreciable $e_{bin}$ growth that is still less than the $e_{bin} = 0.1032$ case (see Section \ref{BinaryEvolution}).  For
the $e_{bin} = 0.25$ case, the binary eccentricity may be large enough that higher order resonances in the disc begin to impact the evolution \citep{Arty92}, potentially accounting for the system's departure from expected behavior.

		%% disc Structure Evolution Subsubsection %%

\subsubsection{Disc Structure} \label{discStructEvolution}

\begin{figure*}
	\includegraphics[width=\textwidth]{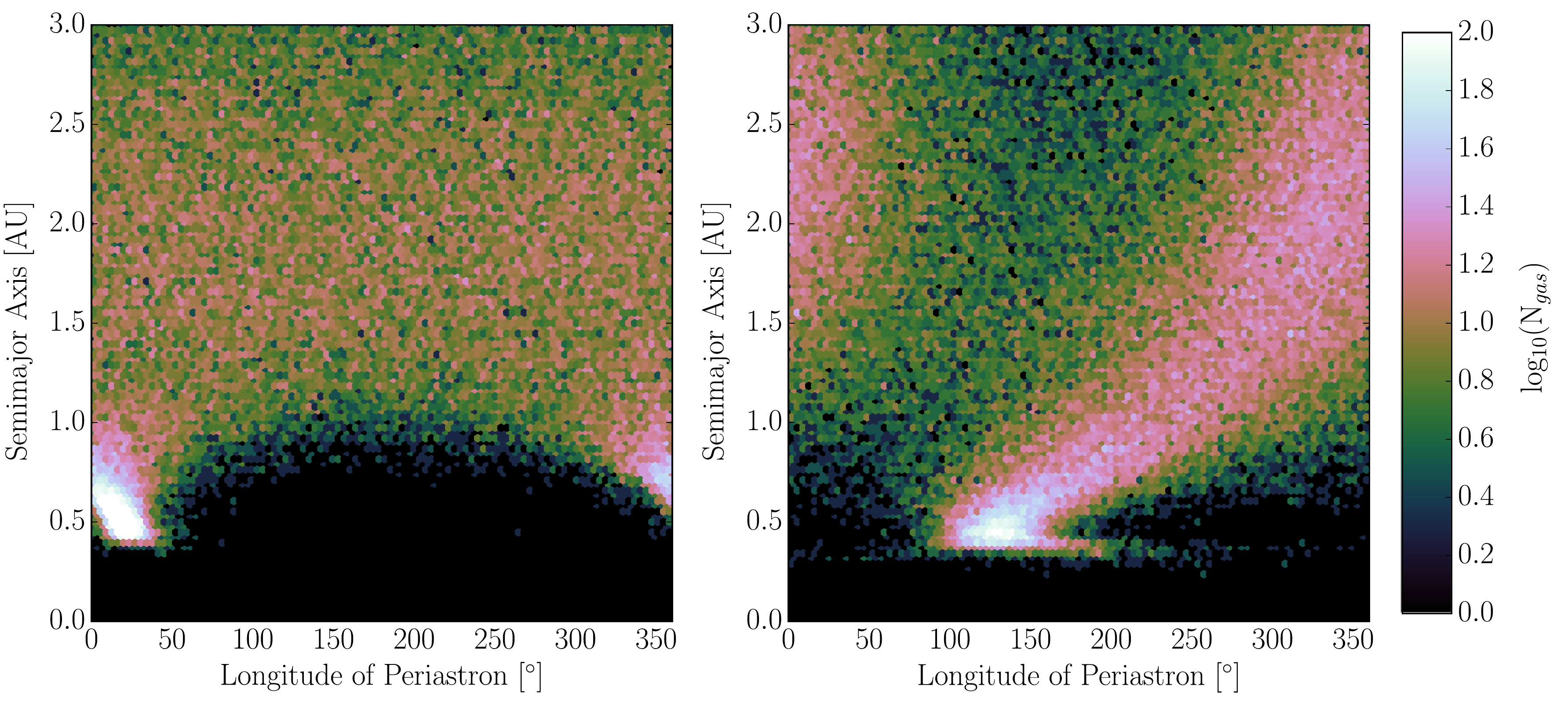}
    \caption{Two dimensional histogram of disc gas particles binned by semimajor axis and
longitude of periastron, $\varpi$, (the sum of the longitude of ascending node, $\Omega$, and argument of periastron, $\omega$)
relative to $\varpi$ of the binary.  Each bin is colored by the log of the number of gas particles it contains.  We note that figure displays the disc in orbital element 
space and not in configuration space.  The left and right panels display the histograms for two simulations with binaries having initial eccentricities of 0 and 
0.1032, respectively, both after 200 years of evolution.  The right panel clearly displays a $m = 1$ spiral wave launched from near the 1:3 EOLR within the disc.}
    \label{fig:figure5}
\end{figure*}

\begin{figure*}
	\includegraphics[width=\textwidth]{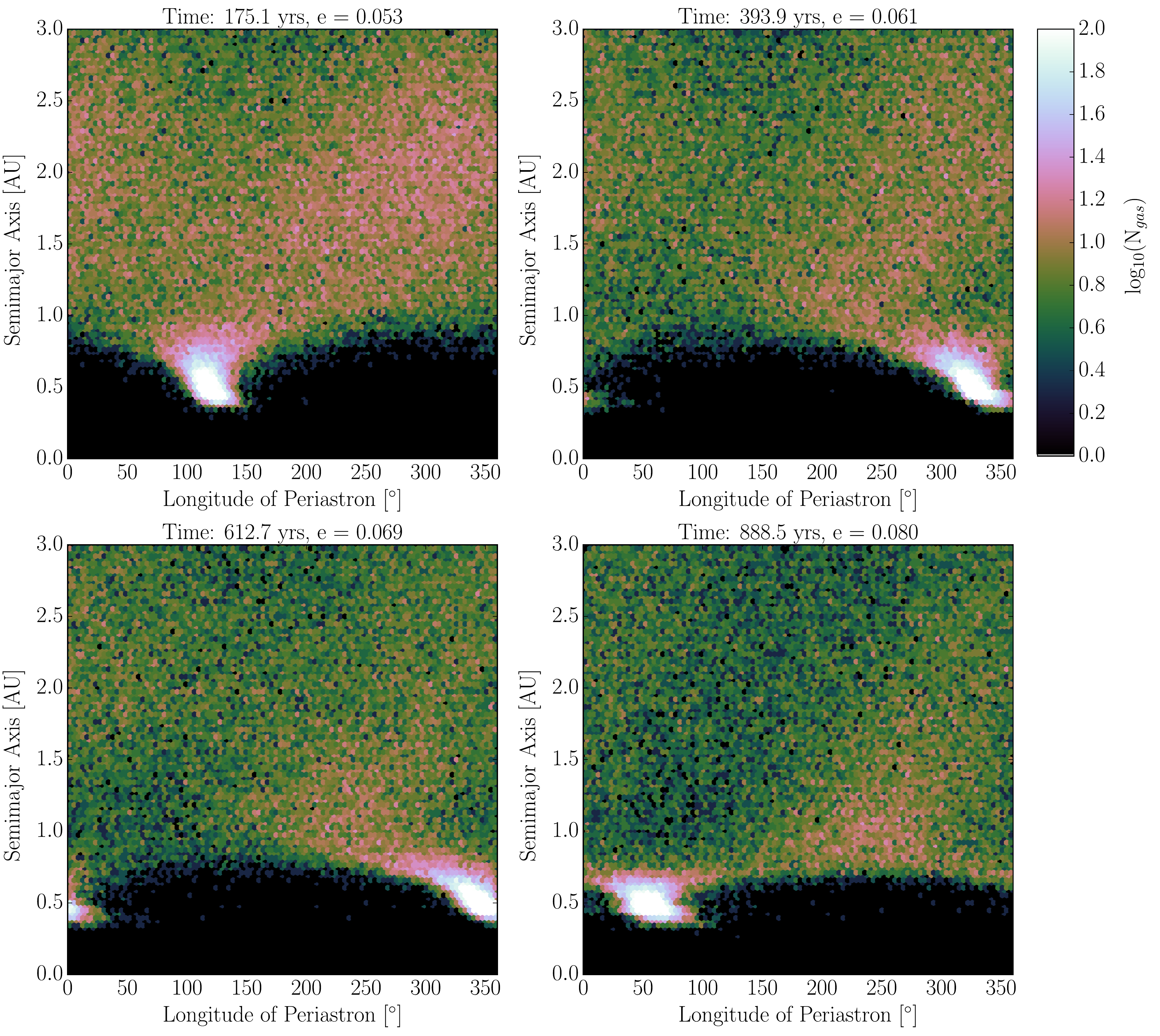} %%% Note: replaced f6 with f6.pdf for the ArXiv
    \caption{Two dimensional histogram of disc gas particles binned by semimajor axis and
longitude of periastron, $\varpi$ of the form of that shown in Fig.~\ref{fig:figure5} for a binary with $e_i$ = 0.05.  We stress that the plot is shown in orbital 
element space and not configuration space.  As the binary eccentricity increases through 
interactions with the surrounding disc, a single weak spiral arm begins to form.  Once $e \approx$ 0.08, the spiral arm becomes slightly more apparent, 
resembling a fainter version of the arm shown in Figure \ref{fig:figure5} for the $e_i$ = 0.1032 case.}
    \label{fig:figure6}
\end{figure*}

The gravitational influence of the binary forces several major changes within the structure of the circumbinary disc.  To explore how the
structure of the disc changes with time and $e_{bin}$, we examined the orbits of gas
particles and how they vary with distance from the binary.  To
accomplish this, two-dimensional histograms of all gas particles
within a radial distance of 3 AU from the barycenter were made for
each snapshot.  We computed the histograms over semimajor
axis, $a$, and the longitude of periastron, $\varpi$, defined in this
work as the sum of the argument of periastron, $\omega$, and the
longitude of the ascending node, $\Omega$, relative to that of the binary, $\varpi_{bin}$, for each gas particle. 
Fig.~\ref{fig:figure5} and Fig.~\ref{fig:figure6} shows these histograms
for all particles in the disc out to 3 AU
for several of our simulations.

In all simulations, a precessing overdense knot was found just outside
inner edge of the disc.  The knot corresponds to a coherent precession
of eccentric gas particle orbits at the inner disc edge.  The knot
depicted in Fig.~\ref{fig:figure5} precesses relative to the binary
in the prograde sense with a period of about 20 years, or 400 binary
periods, for $e_{bin} \approx 0$.
Fig.~\ref{fig:figure5} shows the $a - \varpi$ histogram 
for two simulations with initial $e_{bin}$ of 0 and 0.1032,
respectively after 200 years of evolution.
Precession of orbits near the binary at the inner disc edge are expected due to the binary's time-varying potential as shown in the simulations and analytic theory of \citet{Arty2000}. The existence of this knot is consistent with the identification of a similar overdense lump located at the inner edge of 
gaseous circumbinary accretion discs about binary black holes from 2D simulations by \citet{Farris14}.

An additional structure identified in the disc is a single arm ($m = 1$) spiral wave launched from near the inner edge of the disc as shown
in Fig.~\ref{fig:figure5}.  The spiral wave is an alignment of gas particle longitude of periastrons relative to the binary's.  For the initial $e_{bin} = 0.1032$ simulation, the wave develops rapidly within the first 50 years, or about $10^3$ binary periods.  Recent N-body simulations of circumbinary planetesimal discs by \citet{Lines16} confirmed the presence of an $m = 1$ wave discs about eccentric binaries.  When under the influence of an asymmetric gaseous disc potential, \citet{Lines16} identified the wave as a preferential alignment of planetesimal longitude of periastrons as a function of orbital radius around the eccentric binary of the Kepler 16 system, similar to the spiral wave found in this work.

%%% Figure 5: Disc structure histogram (circular vs eccentric)

In our simulations, the spiral wave's orientation remains locked to
the binary's slow prograde $\varpi_{bin}$ precession throughout the
entire simulation, although a slight drift of a degree or so does
occur.  The wave's fixed orientation relative to the binary is an
important effect whose consequences will be examined more carefully in
Section \ref{BinaryEvolution}.  Simply put, if the wave did circulate
relative to the binary, it's long term effect on the system, if any,
would average out to zero, so the fixed orientation could indeed dynamically impact the system.

Note that in the simulation with an initial $e_{bin} = 0$ (left panel), no spiral $m = 1$ 
arm exists in contrast to the $e_{bin} = 0.1032$ simulation (right panel) which
shows a prominent spiral arm.  The arm is also observed in the
simulation with initial $e_{bin} = 0.25$ but not with initial $e_{bin}
= 0.001$.  Since the spiral arm is only observed when the binary has
an appreciable eccentricity, we can infer that a coupling between binary eccentricity
and the inner disc edge impacts its formation.  

To investigate the role of binary eccentricity in exciting the wave, Simulation 3 was ran with initial $e_{bin} = 0.05$ to see if the spiral wave could be excited with an intermediate $e_{bin}$ between the two regimes identified above.  In the intermediate regime as explained in Section \ref{discEccEvolution}, $e_{bin}$ is not large enough to be strongly coupled to the disc while also not low enough to weakly couple to the disc to drive disc eccentricity.  From the onset of the initial $e_{bin} = 0.05$ simulation, a faint $m = 1$ spiral wave appeared and gradually strengthened as shown in Fig.~\ref{fig:figure6}.  The wave, initially weak, became more apparent after about 900 years.  The $m = 1$ spiral arm in this simulation does not become as pronounced as the one seen in the initial $e_{bin} = 0.1032$ simulation, suggesting that the strength of the arm depends on $e_{bin}$ and supporting the notion that this disc-binary system undergoes an intermediate coupling.  

%%% Figure 6: disc structure histogram vs time for e = 0.05

We again apply the theory of \citet{Papaloizou01} to understand the origin and behavior of the spiral wave.  We know from Section \ref{discEccEvolution} that a nonlinear coupling between non-zero disc eccentricity and the binary's tidal potential excites an $m = 2$ spiral density wave from the 1:3 EOLR that mediates angular momentum transfer in the system.  Also, we have shown that the strength of the coupling between the binary and disc, which depends on $e_{bin}$, determines the magnitude of the disc inner edge eccentricity.  Additional structural changes within the disc proceed via a higher order coupling.  

In \citet{Papaloizou01}, the authors show that the $m=2$ density wave
emitted at the 1:3 EOLR can couple back through the binary tidal potential. This additional coupling produces a time independent $m = 1$ wave and an associated potential.  The extra potential from the $m = 1$ wave can allow for the removal of angular momentum from the system via resonant torques.  The $m = 1$ wave produced via the recoupling mechanism is precisely the $m = 1$ spiral wave identified in this work.  The presence of the $m = 1$ wave was reconfirmed via a Fourier decomposition of the disc surface density.  Since we only observe the $m = 1$ spiral wave in discs around eccentric binaries, we infer that this recoupling mechanism only occurs when the disc and binary are strongly coupled.  \citet{Lines16}'s observation of a $m = 1$ spiral wave present in their simulation of a planetesimal disc surrounding the eccentric Kepler 16 binary support this argument.

Since the orientation of the $m = 1$ wave in our simulations remains locked to that of the binary's, it does not circulate and hence is independent of time.  Also, as we will explore in Section \ref{BinaryEvolution}, the additional potential from the $m = 1$ resonantly torques on the binary causing evolution in its orbital eccentricity.  

Next we examine why the orientation of the spiral wave remains fixed relative to the binary.  We apply the analytic theory for circumbinary orbits of \citet{LeungLee13} to partially explain this effect.  The theory, accurate to first order in $e_{bin}$, decomposes the orbit of a test particle about two stars into a superposition of the circular motion of a guiding center and the radial and vertical epicyclic motion due to the non-axisymmetric components of the binary's potential.  \citet{LeungLee13} give the equations for the precession rate of the argument of periastron, $\omega$, and the longitude of the ascending node, $\Omega$, respectfully, to be
%% Apsidal precession rate equations
\begin{equation}
\label{eqn:Periastron}
\dot{\omega} \approx \frac{3}{4} \frac{m_a m_b}{(m_a + m_b)^2} \left( \frac{a_{bin}}{r} \right)^2
\end{equation}

\begin{equation}
\label{eqn:Longitude}
\dot{\Omega} \approx -\frac{3}{4} \frac{m_a m_b}{(m_a + m_b)^2} \left( \frac{a_{bin}}{r} \right)^2
\end{equation}
where $m_a$ and $m_b$ are the masses of the primary and secondary stars, respectively, $a_{bin}$ is the binary semimajor
axis and $r$ is the radial distance from the barycenter. 

Since these rates are approximately equal and opposite, one expects a
gas particle's longitude of periastron, as defined earlier, to remain
fixed, as is observed for the spiral waves in our simulations about
sufficiently eccentric binaries.  In the context of gaseous circumbinary
discs, this interpretation has a few potential shortcomings.  First, a
given gas particle does not live in isolation since it feels the
effects of disc self gravity and pressure gradients within the disc that can impact its orbit.  Also since this theory is only linear in $e_{bin}$, its applicability could 
lessen as the binary becomes more eccentric due to interactions with the disc.  However, the binaries considered have low to moderate eccentricities and the 
discs are rather low-mass such that disc self gravity is negligible so the gravitational influence of the binary should dominate.  Therefore, we expect this 
theory to still provide a decent approximate explanation for why the spiral arm remains fixed relative to the binary.  
	
		%% Binary Evolution Subsection %%

\subsection{Binary Evolution} \label{BinaryEvolution}

Secular theory \citep{GT79,GT80,Pringle91,Papaloizou01} and previous simulations \citep{Arty91,Cuadra09,Roedig12,Dermine13} show that angular 
momentum losses to a disc change the central binary's orbital elements.  Angular momentum loss occurs mainly through resonant gravitational torques 
at the Lindblad and corotation resonances.  In the case of binary stars embedded in an external disc, loss of angular momentum can result in changes to the 
binary eccentricity and semimajor axis.  For unequal mass binary stars with low to moderate eccentricity, the majority of eccentricity growth is due to resonant 
torques at the 1:3 EOLR.  This resonance dominates since these systems tend to open a gap in the disc such that the 1:3 EOLR resides nearest to the inner 
edge of the disc while the eccentricity damping 1:2 corotation resonance lies within the evacuated region \citep{Arty91,Arty92}.  Simulations of
binaries embedded in circumbinary discs by \citet{Roedig12} agree with this interpretation as they identified gravitational torque density peaks responsible for 
binary eccentricity evolution located at the 1:3 EOLR in the disc.  For the simulations presented in this work, we expect a secular increase in binary eccentricity 
and decrease in semimajor axis as the binary loses angular momentum to circumbinary disc.

%%% Figure 7: Binary semimajor axis vs time
\begin{figure}
	\includegraphics[width=\columnwidth]{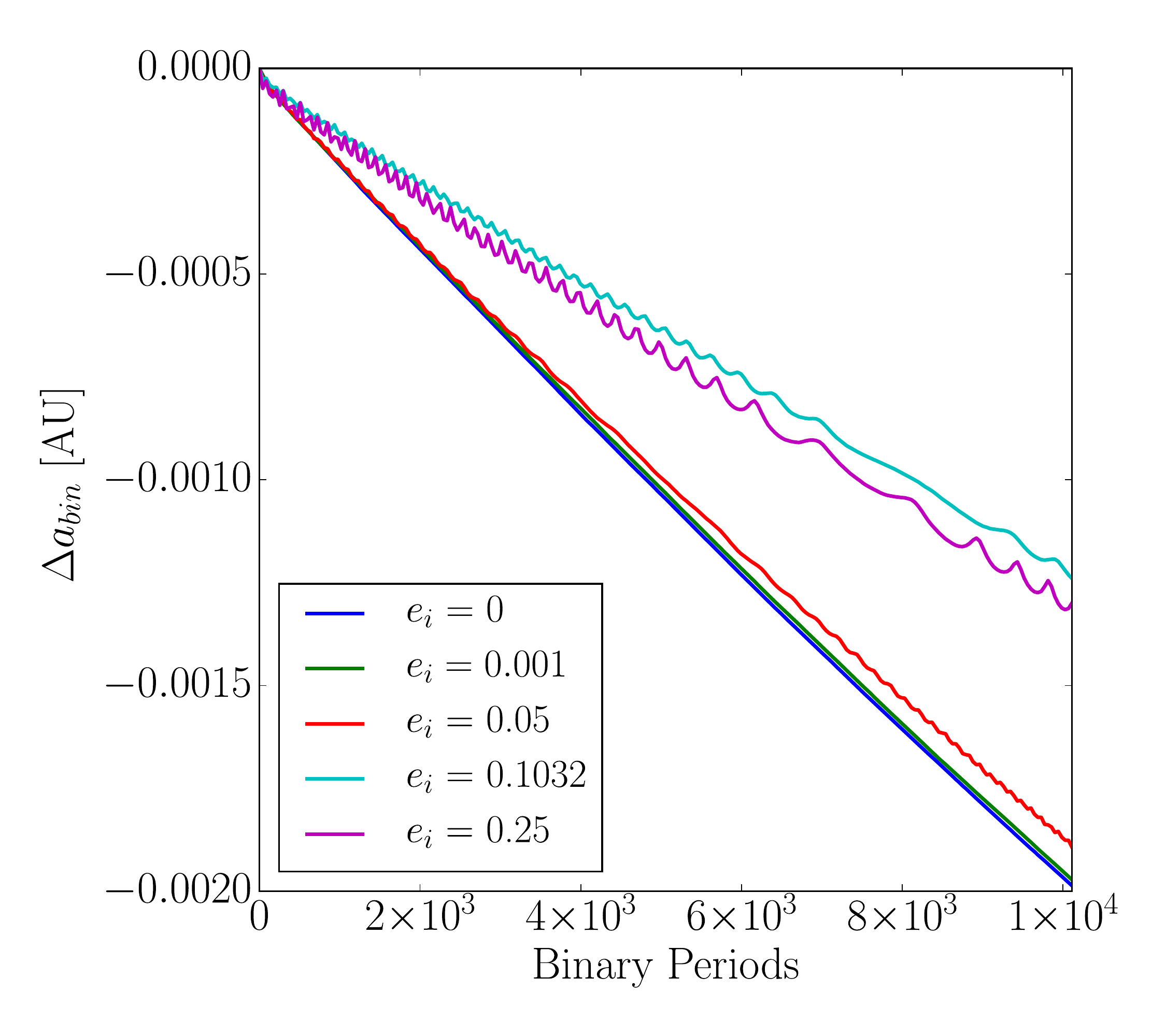}
    \caption{Change in binary semimajor axis versus simulation time in units of binary periods for several initial binary eccentricities. For all cases, the binary's semimajor axis secularly decreases.}
    \label{fig:figure7}
\end{figure}

As shown in Fig.~\ref{fig:figure4}, binary eccentricity for
initially eccentric binaries grows over the duration
of the simulation.  The eccentricity growth rate, $\dot{e}_{bin}$,
seems to scale with initial $e_{bin}$.  In the simulation with initial
$e_{bin} = 0.25$, however, $e_{bin}$ increases more slowly than the
initial $e_{bin} = 0.1032$ case.  When the initial $e_{bin} \approx
0$, no binary eccentricity growth occurs in contrast to the results of
similar simulations of binaries embedded in an external disc by
\citet{Papaloizou01,PierensNelson07,Cuadra09} who find significant
eccentricity growth with an initial $e_{bin} \approx 0$, a discrepancy
we will address later.  In all cases, $e_{bin}$ oscillates as the simulations progress.  This oscillation is due to forcing by the $m = 1$ potential of the eccentric 
external disc \citep{Arty2000}.

For all simulations, the binary semimajor axis secularly decreases where the rate of decline is lower for increasingly eccentric binaries as shown in Fig.~\ref{fig:figure7}.  As before, the initial $e_{bin} = 0.25$ case defies this trend as it shows a greater semimajor axis decline than the initial $e_{bin} = 0.1032$ case instead of the expected lesser decline.  Additionally, the binary's longitude of periastron slowly precesses in the prograde sense less than $1^{\circ}$yr$^{-1}$ over the duration of the simulation similar to the results of comparable simulations by \citet{KleyHag15}.

		%% Subsubsection: Modeling Binary Evolution

\subsubsection{Modeling Binary Evolution} \label{ModelBinaryEvolution}

Any changes in the binary orbital elements will be depend on the
details of the binary's interactions with the disc.  As discussed in
Section \ref{discEccEvolution}, we applied the secular theory of
\citet{Papaloizou01} to show that the strength of the disc-binary
coupling dictates how eccentricity grows within the system.   We apply the same arguments used above to understand disc eccentricity growth to binary eccentricity evolution.  For circular binaries, we argued that the binary and disc are weakly coupled since the mass of the secondary is much greater than the mass of the entire circumbinary disc.  In this weak coupling regime, the eccentricity of the less massive of the binary-disc system grows, which in this case is the disc.  As expected, the disc develops appreciable eccentricity.  From these arguments, we expect the binary to develop very little, if any, eccentricity.  This behavior is exactly what we observe in our simulations (see Fig.~\ref{fig:figure4}).

%%% Figure 8: Fit vs simulation results in e-a plane
\begin{figure}
	\includegraphics[width=\columnwidth]{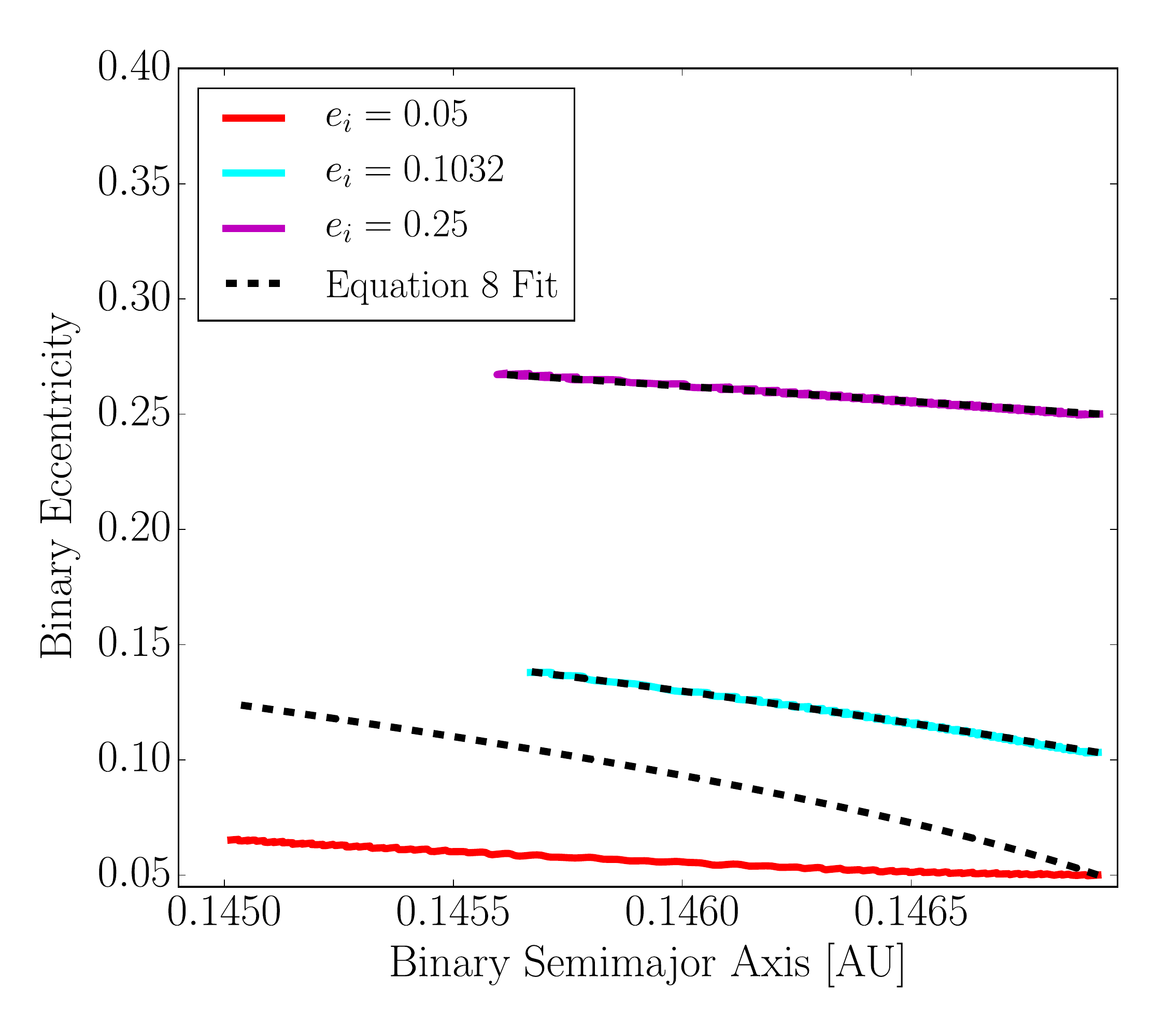}
    \caption{Evolution of $e_{bin}$ as a
      function of $a_{bin}$ from simulations with initial $e_{bin}$ of
      0.05, 0.1032, and 0.25 over-plotted with the analytic fit of
      equation \ref{eqn:total_deda} with $\alpha_{eff} = 0.006$.
      As the system evolves, time advances towards the left in this
      depiction.}
    \label{fig:figure8}
\end{figure}

For simulations of discs around eccentric binaries, we argued that the disc and binary are strongly coupled through the $m = 1$ modes of the eccentric binary orbit and inner disc edge orbits.  In this regime, both the binary and disc eccentricities grow together and further coupling between the disc and binary can occur.  Additional disc-binary coupling discussed at length in Section \ref{discStructEvolution} lead to the excitation of a time-independent $m = 1$ spiral wave and associated potential from the 1:3 EOLR whose orientation remains locked to that of the binary's orbit (see Fig.~\ref{fig:figure5}).  Since this wave remains fixed relative to the binary, it resonantly torques the binary through the 1:3 EOLR, removing angular momentum from the binary's orbit, increasing $e_{bin}$.  If the wave instead circulated over time, its potential would time-average to zero and have no effect on the binary orbital elements.  The presence of the $m = 1$ spiral wave leads to the qualitatively different binary eccentricity evolution for the $e_{bin} = 0.1032$ case relative to the $e_{bin} \approx 0$ case displayed in Fig.~\ref{fig:figure4}.  In all simulations, the binary semimajor axis decreases due to energy dissipation from the viscous disc. 

 Similar N-body SPH simulations of
binaries embedded in a circumbinary disc by \citet{Arty91} showed that for binaries with mass ratio $\mu = m_2/m_1 = 0.3$ and $e_{bin} \approx 0.1$ similar to the systems examined in this paper, resonant interactions with the surrounding disc at the 1:3 EOLR drive eccentricity growth and semimajor axis decay.  In this regime, we expect the binary eccentricity growth observed by \citet{Arty91} since the binary and disc are strongly coupled.

With the origin of binary eccentricity evolution understood using
the theory of \citet{Papaloizou01}, we seek to quantify binary orbital
evolution.  Following the analysis of \citet{Dermine13}, we applied the theory of \citet{Arty96b,Arty2000} to quantify the eccentricity and semimajor axis 
evolution of a binary with initial $e_{bin} \approx 0.1$ embedded in an external gaseous disc due to resonant interactions with the 1:3 EOLR using the following 
relation
%% de/da equation from Artymowicz+Lubow  2000
\begin{equation} 
\label{eqn:deda}
\dot{e} = \frac{1 - e^2}{e} \left(\frac{l}{m} - \frac{1}{\sqrt{1 - e^2}}\right)\frac{\dot{a}}{a}.
\end{equation}
where $(l,m) = (1,2)$ is the potential component corresponding to the 1:3 EOLR \citep{Arty91,Arty2000} and $e$ and $a$ are the binary eccentricity and semimajor axis, respectively.

For less eccentric binaries, the semimajor axis and eccentricty evolution is well-described by 
%% de/da equation from Artymowicz+Lubow 1996
\begin{equation}
\label{eqn:deda1996}
\dot{e} = -\frac{50e}{\alpha_{eff}}\frac{\dot{a}}{a}
\end{equation}
where $\alpha_{eff}$ is the effective standard viscosity parameter \citep{Arty96b,Dermine13}.  

Note that $\alpha_{eff}$ in equation \ref{eqn:deda1996} is not in 
general the same as the $\alpha_{SPH}$ viscosity parameter discussed in Section
\ref{methods_section}.  To relate $\alpha_{eff}$ and $\alpha_{SPH}$, we use the following relation from \citet{Lodato10} and \citet{Meru12}

%% alpha_eff <--> alpha_SPH relation 
\begin{equation}
\label{eqn:alphas}
\alpha_{eff} = \frac{k_{BSW}}{20} \alpha_{SPH} \frac{h}{H}
\end{equation}
where $h$ is the smoothing length, $H$ is the disc aspect ratio and the factor of $1/20$ comes from the \citet{Meru12} derivation of the coefficient for 
the \citet{Monaghan83} viscosity implementation used in ChaNGa \citep{Murray96}.  The $k_{BSW}$ factor arises from our use of the  Balsara switch which 
limits shear viscosity by scaling both $\alpha_{SPH}$ and $\beta_{SPH}$ \citep{Balsara95}.  The range of $k_{BSW}$ is [0,1].  Since both $h$ and $k_{BSW}$ 
can vary between gas particles and $H$ can vary radially as the disc evolves, we average over the disc to get $k_{BSW} = 0.3$ and $h/H$ = 0.4.  Given these 
values, we set $\alpha_{eff} = 0.006$ as the approximate value for our simulations with $N_{gas} = 10^5$.

Equations \ref{eqn:deda} and \ref{eqn:deda1996}, derived by \citet{Arty96b} via examining the balance between viscous and resonant interactions at the inner 
disc edge, apply in separate regimes that depend sensitively on $e_{bin}$.  \citet{Arty96b} estimate that once $e_{bin} \gtrsim 0.1 \alpha_{eff}^{1/2}$, the 1:3 
EOLR dominates binary eccentricity growth while below this threshold for circular binaries, no eccentricity growth occurs.  These separate regimes correspond 
to the weak and strong disc-binary coupling for circular and eccentric binaries, respectively, discussed in previous sections.  For nearly circular binaries, 
$e_{bin} < 0.1 \alpha_{eff}^{1/2}$ and $\dot{e}_{bin} \approx 0$ as expected for weak disc-binary coupling.  Conversely for eccentric binaries, $e_{bin} \gtrsim 
0.1 \alpha_{eff}^{1/2}$ and $\dot{e}_{bin} > 0$ as demonstrated above for strong disc-binary coupling.  We combine equations \ref{eqn:deda1996} and 
\ref{eqn:deda} to model how $e_{bin}$ and $a_{bin}$ should evolve under the influence of an external disc following the models of \citet{Arty96b,Arty2000} and 
\citet{Dermine13}
%% Total de/da model from Artymowicz+Lubow 1996b,2000 similar to Dermine 2013
\begin{equation}
\label{eqn:total_deda}
de/da=
\begin{cases}
-\frac{50e}{\alpha_{eff}}\frac{1}{a} & \text{if } e_{bin} \lesssim 0.1 \alpha_{eff}^{1/2}\\
\frac{1 - e^2}{ea} \left(\frac{l}{m} - \frac{1}{\sqrt{1 - e^2}}\right) & \text{if } e_{bin} \gtrsim 0.1 \alpha_{eff}^{1/2}.
\end{cases}
\end{equation}
Hence, eccentricity increases while the semimajor axis decreases.

To verify that equation \ref{eqn:total_deda} is a proper model for our simulations, the observed evolution of $e_{bin}$ and $a_{bin}$ for simulations with initial $e_{bin} =$ 0.05, 0.1032, and 0.25 were compared with the theoretical result of equation \ref{eqn:total_deda} assuming $\alpha_{eff} = 0.006$.  The comparison is shown in Fig.~\ref{fig:figure8}.

The results of both the simulations with initial $e_{bin} = 0.1032$ and $e_{bin} = 0.25$ are in good agreement with the theoretical expectations of equation \ref{eqn:deda} and also in accordance with the simulation of a similar system with initial $e_{bin} = 0.1$ by \citet{Arty91}.  Equation \ref{eqn:total_deda} can also be applied to the initial $e_{bin} \approx 0$ cases.  For these simulations, the binary semimajor axis decreases via energy dissipation through the viscous disc while no eccentricity growth occurs as seen in Fig.~\ref{fig:figure4} and Fig.~\ref{fig:figure7}.  Our results are consistent with the prediction of equation \ref{eqn:total_deda}.  These findings indicate that equation \ref{eqn:total_deda} successfully quantifies how the binary evolves in the different disc-binary coupling regimes.

The initial $e_{bin}$ = 0.05 case proves troublesome.  Equation \ref{eqn:total_deda} does a poor job fitting the binary eccentricity and semimajor axis evolution.  The poor fit can be understood in the context of how the disc and binary undergo an intermediate coupling in between the strong and weak regimes.  As discussed in Sections \ref{discEccEvolution} and \ref{discStructEvolution}, the binary eccentricity is not large enough to launch a prominent $m = 1$ spiral wave and drive the eccentricity growth for more eccentric binaries.  In between regimes, we expect the binary eccentricity to grow weakly as we observe in our simulations.  Equation \ref{eqn:total_deda} succeeds for systems firmly in the weakly or strongly coupled regime but does not perform well for intermediate coupling.

This behavior has interesting consequences for the subsequent evolution of a system.  For the intermediate case as the binary eccentricity grows with time, it will eventually reach $e_{bin} \approx 0.1$ and will then begin to strongly couple to the gaseous disc.  As discussed previously, strong coupling launches a $m = 1$ spiral wave in the disc and increases the growth rate of binary eccentricity.  Also in the strongly coupled regime, we expect the disc eccentricity to decrease from higher values and be similar in magnitude to the binary eccentricity.  

In a similar vein for nearly circular binaries, we expect them to remain circular.  Other simulations of binaries embedded in circumbinary discs such as those by \citet{PierensNelson07} and \citet{Cuadra09} have found that initially circular binaries eventually develop appreciable non-zero eccentricities.  This behavior can be understood by examining equation \ref{eqn:total_deda}.  If the binary is perturbed and some non-zero eccentricity develops, we would expect the binary eccentricity to grow very slowly, gradually strengthening the coupling between the disc and binary until intermediate coupling is reached and the system progresses as described above.  This picture is consistent with the results of \citet{PierensNelson07} who found the binary eccentricity began to grow on timescales longer than those explored in this work.  Therefore over timescales much longer than simulated here, we would expect our initial $e_{bin} = 0.01$ case to become appreciably eccentric.  We extrapolate the results of our simulations to longer timescales and consider the consequences below.

We ran additional simulations to ensure that  the 1:3 EOLR did indeed dominate binary
evolution and no other effect played a major role.  To do this, a shorter simulation with an initial binary eccentricity of 0.1032 was performed
with the initial disc gap radius located outside of the 1:3 EOLR.
Minimal binary eccentricity growth and effectively no binary semimajor
axis decay occurred until the disc viscously spread inward.  Additionally, the $m = 1$ spiral wave observed in other simulations of sufficiently eccentric binaries also did not exist until mass was able to drift inward and accumulate at the 1:3 EOLR at which point the $e_{bin}$ and $a_{bin}$ began to evolve.  These findings support the supposition that interactions with the 1:3 EOLR drives the binary evolution as anticipated.

To study what effect, if any, accretion has on the how the binary stars' orbital elements vary, a procedure similar to that used by \citet{Roedig12} 
was performed.  For a given simulation each accretion event was tracked such that the accreted gas particle's mass and velocity components were
outputted.  Using these events, $e_{bin}$ and $a_{bin}$ were evolved by adding each accreted particle to the binary imposing linear
momentum and mass conservation as is done natively in ChaNGa for sink particles.  This test demonstrated that accretion 
had a negligible effect on $e_{bin}$ and $a_{bin}$.  Therefore, it is safe to assume that the evolution of the binary orbital elements is primarily driven by interactions with the external disc, in particular at the 1:3 EOLR for the systems considered in this work.

%% Effects of varying disc properties %%

\section{Effects of Varying Disc Properties} \label{VaryingDiscProps}

Previous studies have examined how varying disc properties can change
how a circumbinary disc evolves.  For example, \citet{Lines15} showed
that disc eccentricity is sensitive to the initial disc surface
density gradient and aspect ratio.  Here, we analyse the results of simulations that vary disc mass, gas resolution, and
aspect ratio in order to examine how disc properties impact the disc - binary coevolution.

	%% Varying disc mass

\subsection{Varying Disc Mass} \label{VaryingDiscMass}

To study how our results vary with disc mass, three additional simulations were run with $e_{bin} =
0.1032$ to see if varying disc mass changes binary evolution.  Simulation 6
with $2 \times M_{disc}$, Simulation 7 with $0.5 \times M_{disc}$, and
Simulation 8 with $1.5 \times M_{disc}$ were run.  For these additional simulations, we expect the disc to be strongly coupled to the eccentric binary since the 
coupling only depends on the magnitude of binary eccentricity.  We do expect, however, that eccentricity growth to occur more quickly for systems with more 
massive discs since torque scales with the disc mass.  The results of the simulations are shown in Fig.~\ref{fig:figure9}.

The disc eccentricity of Simulation 7 was similar to that of Simulation 4 while Simulation 6 showed larger disc eccentricity values.  As expected, more massive 
discs became more eccentric.  Not depicted is the spiral arm development.  Similar to the standard initial $e_{bin} = 0.1032$ case, a prominent $m = 1$ spiral 
wave quickly forms in Simulations 6, 7, and 8 consistent with the picture that binary eccentricity determines how strongly the disc and binary couple.  In all three 
simulations, the wave has the same shape and remains fixed relative to the binary except for Simulation 6 which showed slight prograde precession of the 
spiral arm.  

For more massive discs, binary eccentricity grew more quickly and the binary underwent more semimajor axis decay.  Although not plotted, both Simulations 6, 7 and 8 are still well-described by equation \ref{eqn:total_deda} and hence correspond to
either faster or slower binary evolution timescales.  

	%% Varying disc resolution

\subsection{Varying Disc Resolution} \label{VaryingDiscResolution}

To ensure that our simulations were sufficiently resolved, we ran two additional simulations, Simulations 9 and 10, which has decreased and increased
 the initial number of gas particles by a factor of 2 to $5 \times 10^4$ and $2 \times 10^5$, respectively.  The results of these simulations are shown in Fig.~\ref{fig:figure9}.
 
 In both simulations, we find the general trend of eccentricity growth and semimajor axis decay holds.  The lower resolution Simulation 9 eccentricity 
 growth is less than the fiducial Simulation 4.  In addition, the disc eccentricity does not oscillate as seen in other simulations.  This suggests that the clump 
 which dominates the disc eccentricity does not form into a coherent structure, indicating that N$_{gas} = 5 \times 
 10^4$ might not be large enough to properly resolve all the physics at the disc inner edge.  We find that the $m = 1$ spiral wave appears in the disc indicating 
 that the binary strongly couples to the disc as we expect from our previous simulations.
 
 The higher resolution Simulation 10 exhibits both binary and disc eccentricity evolution that is in good agreement with the fiducial Simulation 4.  
 Additionally, we again observed a prominent $m = 1$ spiral wave within the disc that behaved identically to its Simulation 4 counterpart.  One disagreement 
 between Simulation 10 and Simulation 4 is that the higher resolution simulation displayed less binary semimajor axis decay.  This result is expected, 
 however, since higher resolution N-body SPH simulations will have smaller gas softening lengths, $h$.  As shown in equation \ref{eqn:alphas}, the effective 
 standard viscosity parameter $\alpha_{eff} \propto h$.  The theoretical work of \citet{Arty96b,Arty2000} estimate that $\dot{a}/a 
 \propto -\alpha_{eff}$.  We therefore expect a higher resolution run with smaller $h$ and hence smaller $\alpha_{eff}$ to exhibit less binary semimajor axis 
 decay due to dissipation from the viscous disc.
 
Since the main effects explored in this work, the binary eccentricity evolution and the accompanying development of disc eccentricity and structure, are 
in good agreement between the standard and higher resolution runs, we find that our nominal resolution of N$_{gas} = 10^5$ is sufficient.

	%% Varying disc aspect ratio
	
\subsection{Varying Disc Aspect Ratio}

Simulations of accretion discs around binary black holes have examined the effects of larger aspect ratio discs, mainly focusing on accretion rates.  
Two dimensional SPH simulations of gas accretion onto binaries embedded in a circumbinary disc by \citet{Young15} found that increasing gas temperature 
leads to increased accretion rates onto the primary star and growth in the binary mass ratio.  Simulations of discs about black hole binaries by \citet{Ragusa16} 
showed that discs with aspect ratios H/R $\gtrsim 0.1$ have enhanced accretion rates as the inspiralling gas is not supressed by the binary's gravitational 
torque.  To examine disc aspect ratio's effect on disc-binary evolution in our simulations, we ran Simulation 11 with a disc aspect ratio of $H/R = 0.12$.  To 
initialize the disc with a larger aspect ratio, we increased the disc temperature by setting T$_0$ in equation \ref{eqn:disc_temp_profile} to 2500 K giving us 
about a factor of $2$ larger aspect ratio relative to the fiducial Simulation 4.

\begin{figure*}
	\includegraphics[width=\textwidth]{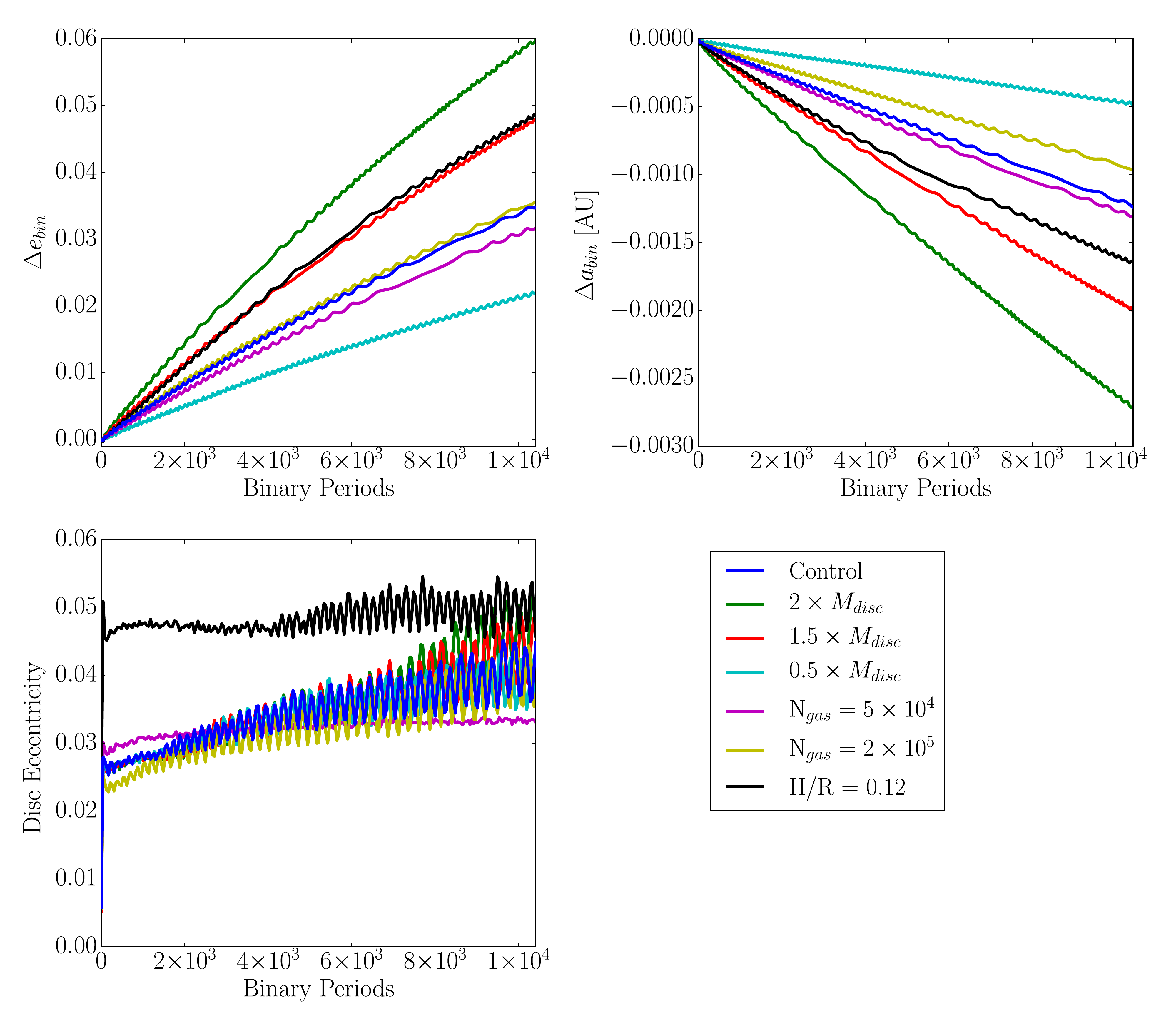}
    \caption{The results of simulations about a binary with initial $e_{bin} = 0.1032$ and $a_{bin} = 0.1469$ AU with various disc properties.  The control case corresponds to Simulation 4.  The top left panel gives the change in binary eccentricity as a function of time.  The top right panel shows the change in binary semimajor axis vs time.  The bottom left panel displays disk eccentricity vs time and the bottom right panel displays the figure legend.}
    \label{fig:figure9}
\end{figure*}

The result of Simulation 11 is presented in Fig.~\ref{fig:figure9}.  We found that the thicker disc resulted in greater binary semimajor axis decay and 
eccentricity growth compared to the fiducial Simulation 4.  The increased binary semimajor axis decay agrees with the theoretical expectation of 
\citet{Arty96b,Arty2000} who estimate that $\dot{a}/a \propto -(H/R)^2$.  The enhanced binary eccentricity growth follows as a consequence of equation 
\ref{eqn:deda}.

In Simulation 11, the binary accretion rate was enhanced by about a factor of $4$ relative to the fiducial Simulation 4 in agreement with the general 
findings of both \citet{Young15} and \citet{Ragusa16}.  We found that accretion had little impact on the binary orbital element evolution.  As expected from arguments presented in Section \ref{discStructEvolution}, the disc and binary were strongly coupled, producing a prominent $m = 1$ spiral wave similar to the one seen in Fig.~\ref{fig:figure5}.  The disc also displayed larger eccentricity initially but it did not grow appreciably over the course of the 
simulation.  
	
%%% Figure 9: Effects of varying disc props
	
%% Discussion Section %%

\section{Discussion} \label{discuss}
The coevolution of a binary with a gaseous circumbinary disc, primarily driven by resonant interactions at the 1:3 EOLR, has several important consequences for the subsequent dynamical evolution of the system.  As shown above, an eccentric binary system tends to gain eccentricity and experience a secular decay in semimajor axis due to viscous and resonant interactions with the disc.  This evolution not only changes due to additional feedback with the disc, but also impacts regions in the disc where planets form and migrate.

	%% Subsection: Potential tidal effects/coalescence 
	
\subsection{Implications for Long-term Binary Evolution} \label{tidal_implications}

N-body simulations of an unequal mass binary embedded in a
protoplanetary disc by \citet{Arty91} found rapid semimajor axis decay leading the
authors to suggest that the binary separation may become small enough that tidal effects or even stellar coalescence may occur for such systems.  Tides 
between stellar companions tend to circularize the orbit over long timescales once the stellar separation becomes sufficiently small.  Detailed studies of 
companions to Sun-like stars by \citet{Raghavan10} and measurements of solar-type spectroscopic binaries in M35 by \citet{Meibom05} both found that 
binaries from these populations with periods less than about 10 days tend to be circularized.  Theoretical work on the premain-sequence evolution of 
$0.5-1.25$ $M_{\odot}$ binaries by \citet{Zahn89} demonstrate that binaries with orbital periods of about 8 days or less are tidally circularized with effectively all 
of the circularization occurring before the stars reach the main-sequence.  For binaries with an initial period slightly greater than the $\sim 10$ day tidal 
circularization boundary embedded in a circumbinary disc, binary-disc interactions could potentially decrease $a_{bin}$ enough to make tidal effects important 
for subsequent evolution given that the lifetimes of protoplanetary discs are of order 1 Myr \citep{Haisch01}.

In addition, we would expect some longer period binaries to develop appreciable eccentricity through this mechanism.  Observations of spectroscopic binaries discussed by \citet{Mazeh08} show that a large number of such binaries have large eccentricities, some up to $e_{bin} \approx 1$, suggesting that disc-binary interactions may in fact be an important mechanism in pumping binary eccentricity.  Ideally, additional observations of binaries with circumbinary planets, systems guaranteed to have had protoplanetary discs, will allow us to better constrain and model this effect.

The extent to which disc-binary interactions impact astrophysical systems over the disc's lifetime is difficult to measure.  Over the course of the disc's 
lifetime, what may occur is some process that removes the 1:3 EOLR from the disc.  As $a_{bin}$ decays through disc-binary interactions, the location of the 
1:3 EOLR shifts inwards.  Also as $e_{bin}$ grows, the central gap size increases \citep{Arty94}.  The combined $a_{bin}$ and $e_{bin}$ evolution could result 
in the 1:3 EOLR moving into the evacuated disc gap, removing its influence from the system, leaving higher order resonances to influence the binary.  For 
binaries with large $e_{bin}$, \citet{Arty91} speculates that the combination of higher order inner and outer Lindblad resonances and corotation resonances 
should combine to reduce the magnitude of $\dot{e}_{bin}$ and $\dot{a}_{bin}$, potentially preventing subsequent evolution.  This picture is not so 
simple, however, as simulations of binary SMBHs embedded in gaseous discs by \citet{Cuadra09} and \citet{Roedig11} both find that binary eccentricity growth 
continues to $e_{bin} > 0.35$ where this growth did not slow until $e_{bin} \approx 0.6-0.8$.  We note that the simulations of \citet{Roedig11} assumed a fixed 
$a_{bin}$ which neglects the inward motion of the resonances as $a_{bin}$ decays, potentially leaving them in the evacuated region, removing their effects 
from the system.  The impact of higher order resonances on binary evolution is a complicated matter that requires proper treatment in which both the binary and 
disc are allowed to coevolve together and likely depends on disc structure and artificial viscosity implementation.  Additionally, findings by \citet{Pringle91} show 
that in principle, there is no limit to the amount of angular momentum that can be lost by a central binary to an external disc suggesting that binary coalescence 
is not as unrealistic as it sounds.   We caution that when performing simulations of binaries embedded in a gaseous disc that explore the role of semimajor axis decay, one should ensure that their observed semimajor axis decay has converged as both resolution and non-trivial effects such as accretion \citep[e.g.][]{Roedig12} can have a substantial impact.

One effect not explored in this work is the possibility of Kozai-Lidov (KL) oscillations for the general case of a misalign discs in binary systems.   
For an inclined test particle orbiting one component of a binary, periodic KL oscillations 
allow for the particle's eccentricity to grow at the expense of its inclination \citet{Kozai62,Lidov62}.  For the case of an inclined circumstellar disc about one 
component of the binary, \citet{Martin14} found that the disc can exhibit KL cycles with the periodic disc eccentricity maxima approaching $\sim 0.6$.  A 
later study of similar systems by \citet{Lubow15} demonstrated that misaligned discs can become much more extended than coplanar discs and 
potentially could overflow the Roche lobe of the star.  Simulations of misaligned circumbinary discs by \citet{Nixon13} showed that discs of almost all 
inclinations can tear leading to massive accretion and potentially a merger of the central binary.

Given these results in the general case of systems with 
misaligned circumbinary discs, the binary eccentricity evolution is likely significantly impacted by the disc evolution and depart from the results presented here 
for thin, coplanar discs.  The disc, if misaligned, could reach large eccentricities due to KL oscillations and via interactions with the binary if it does not tear.  If 
the disc does in fact tear, the binary would likely not couple with the disc at all but could in fact coalesce as demonstrated by \citet{Nixon13}.  The general 
case of a binary coupling with an inclined circumbinary disc is greatly complicated by KL oscillations, torque scaling with inclination and the potential for tearing 
and warrants a more robust future study.

	%% Subsections: Implications for CBPs
\subsection{Implications for Circumbinary Planets} \label{cbp_implications}

The observed orbital elements of binary stars that host a circumbinary planet are the product of a complex evolutionary history.  From Fig.~\ref{fig:figure4} and 
Fig.~\ref{fig:figure7}, we see that for systems similar to the ones considered in this work, appreciable changes can occur on order $10^4$ binary orbits.  

As shown in Section \ref{VaryingDiscMass}, the mass of the disc strongly influences the binary evolution.  More massive discs, for example, drive much faster 
$e_{bin}$ growth and $a_{bin}$ decay.  Faster dynamical binary evolution due to massive discs could be particularly relevant for {\em Kepler} circumbinary 
planets as the work of \citet{Dunhill13} suggests that these circumbinary planets formed and migrated in massive discs.  Additionally, disc-binary interactions 
can make planet formation more difficult.  Simulations by \citet{Lines16} identified an $m = 1$ spiral wave in the circumbinary disc that corresponds to an 
alignment of planetesimal longitudes of periastron.  This wave, whose origin was explained in this work, caused an increase in erosive planetesimal 
collisions making in-situ formation difficult in circumbinary protoplanetary discs.

The decay of $a_{bin}$ via disc-binary interactions also causes the inward shift of mean motion and Lindblad resonances.  These resonances can significantly 
impact the orbital stability of local objects in the disc in several important ways.  For the restricted three body problem, resonance overlapping can lead to 
stochastic orbital evolution as shown from the criterion derived by \citet{Wisdom80}.  For the case of binary orbital evolution driven by tides, \citet{Bromley15} 
point out that evolving binary eccentricity and semimajor axis changes the location of critical resonances and hence where they overlap, potentially making 
stable systems unstable over time.  The location of mean motion resonances also dictate where circumbinary planets may reside.  The numerical integrations 
of both \citet{Popova13} and \citet{Chavez15} show that many circumbinary planets lie in a stable region shepherded by unstable mean motion resonances.  If 
$a_{bin}$ evolves significantly on short enough timescales, so too do the locations of the resonances, sweeping inward and potentially destabilizing orbits.  
We note, however, that $a_{bin}$ evolution appears to be a resolution dependent effect which future work should address.

This behavior is of particular importance for studies of planetary migration in circumbinary discs.  Studies of circumbinary planetary migration in a viscous, 
eccentric disc find that planets tend to migrate inwards until they are trapped in or near the 4:1 mean motion resonance in the region of stability identified by 
\citet{HolmanWiegert99} \citep{Nelson03,Kley14}.  Since the resonances and the region of stability move as binary eccentricity and semimajor axis evolve, the 
final location and stability of migrating planets in circumbinary discs is sensitive to binary evolution.  Simulations of circumbinary systems, especially those 
using N-body SPH methods like the ones presented in this work, must ensure that they properly account for the disc-binary interactions.

%% Conclusions %%
\section{Conclusions}

In this work, we showed that unequal mass binary stars embedded in a circumbinary gaseous disc carved out a gap in the disc and caused structural changes 
within the disc.  Resonant interactions with the binary at the 1:3 EOLR excited disc eccentricity.  Sufficiently eccentric binaries excited a $m = 1$ spiral 
wave within the disc.  This wave corresponded to an alignment of gas particle longitude of periastrons that varied with radius.  The spiral wave formed within 50 
years for discs about sufficiently eccentric binaries but took longer to strengthen for less eccentric binaries (see Fig.~\ref{fig:figure5}).  Eccentric binary stars 
became more eccentric and experienced a secular decrease in semimajor axis while initially nearly circular binaries underwent no eccentricity growth over the 
timescales considered.

Eccentricity growth within the system was understood in the context of the theory of \citet{Papaloizou01} in which nonlinear coupling between non-zero disc 
eccentricity and the binary's tidal potential excites an $m = 2$ spiral density wave from the 1:3 EOLR that mediates angular momentum transfer in the system.  
Nearly circular binaries weakly couple to the external disc and drive the inner disc edge to become very eccentric.  Eccentric binaries, however, strongly couple 
to the disc leading to eccentricity growth for both the disc and binary.  The origin of the $m = 1$ wave within the disc is understood as a recoupling of the $m = 
2$ spiral density wave with the binary tidal potential.

This model does have limited applicability as disc gap size scales with $e_{bin}$, so the 1:3 EOLR could fall within the evacuated region removing its effect 
from the system, potentially slowing down binary evolution. For simulations of gaseous circumbinary discs, we caution that the disc-binary interaction must be 
sufficiently accounted for to properly model the system.  We leave the characterization of the long-term impact of disc-binary interactions to future work.

Limitations of this work include the difficulty in integrating the binary orbit.  Since the binary feels the force of every other SPH particle in our simulations and is 
integrated using ChaNGa's native leapfrog integrator, very conservative timestepping was employed to ensure that the binary orbit was well-resolved and 
physically accurate.  The conservative timestepping scheme significantly slowed our simulations.  In light of this limitation, potential future work could include 
running a long-term higher resolution simulation over at least $10^5$ binary orbits for small yet non-zero $e_{bin}$ in order to better 
characterize how the disc and binary coevolve.  Additional future work could involve examining equal mass binaries or binaries with larger eccentricities than 
those explored in this work.  Since binaries with large eccentricities excite higher order resonances within the disc \citep[e.g.][]{Arty92,Arty2000} and carve out 
gaps that could remove the 1:3 EOLR from the disc \citep{Arty94}, it would be interesting to examine how these other resonances can impact binary evolution.  
A study on how different numerical viscosity implementations impact binary evolution would also prove fruitful to examine its influence on disc-binary 
coevolution, specifically binary semimajor axis decay.

\section*{Acknowledgements}

We thank the anonymous referee for helpful comments and suggestions that improved the quality
of the manuscript. We would also like to thank Isaac Backus for a careful reading of the manuscript and
Jacob Lustig-Yaeger for helpful feedback. 
This work was facilitated through the use of advanced computational, storage, and networking
infrastructure provided by the Hyak supercomputer system at the
University of Washington.  We made use of {\em pynbody}
(https://github.com/pynbody/pynbody) in our analysis for this paper.
This work was performed as part of the NASA Astrobiology Institute's
Virtual Planetary Laboratory, supported by the National Aeronautics
and Space Administration through the NASA Astrobiology Institute under
solicitation NNH12ZDA002C and Cooperative Agreement Number
NNA13AA93A.  David Fleming is supported by an NSF IGERT DGE-1258485 fellowship.
Thomas Quinn is supported by NASA grant NNX15AE18G.

%%%%%%%%%%%%%%%%%%%%%%%%%%%%%%%%%%%%%%%%%%%%%%%%%%


\bsp	% typesetting comment
\label{lastpage}

\begin{thebibliography}{99}

%% Aly et al 2015
\bibitem[Aly et al.(2015)]{Aly15} Aly, H., Dehnen, W., Nixon, C., \& King, A.\ 2015, \mnras, 449, 65

%% Armitage 2005: CB black holes

\bibitem[Armitage \& Natarajan(2005)]{Armitage05} Armitage, P.~J., \& Natarajan, P.\ 2005, \apj, 634, 921

%% Artymowicz 1991: Orbital elements changes via external disc

\bibitem[Artymowicz et al.(1991)]{Arty91} Artymowicz, P., 
Clarke, C.~J., Lubow, S.~H., \& Pringle, J.~E.\ 1991, \apjl, 370, L35

%% Artymowicz 1992: Eccentricity changes

\bibitem[Artymowicz(1992)]{Arty92} Artymowicz, P.\ 1992, \pasp, 104, 769

%% Artymowicz and Lubow 1994: Gap sizes

\bibitem[Artymowicz \& Lubow(1994)]{Arty94} Artymowicz, P., \& Lubow, S.~H.\ 1994, \apj, 421, 651

%% Arty 1996a: Interactions of young binaries with protostellar discs
\bibitem[Artymowicz \& Lubow(1996)]{Arty96a} Artymowicz, P., \& Lubow, S.~H.\ 1996, Disks and Outflows Around Young Stars, 465, 115

%% diskpy citation

\bibitem[Backus \& Quinn (2016)]{diskpy} Backus, I., \& Quinn, T.\ 2016, submitted to \mnras

%% Balsara 1995: The balsara switch

\bibitem[Balsara(1995)]{Balsara95} Balsara, D.~S.\ 1995, Journal of Computational Physics, 121, 357

%% Bromley 2015

\bibitem[Bromley \& Kenyon(2015)]{Bromley15} Bromley, B.~C., \& Kenyon, S.~J.\ 2015, \apj, 806, 98

%% Chavez 15

\bibitem[Chavez et al.(2015)]{Chavez15} Chavez, C.~E., Georgakarakos, N., Prodan, S., et al.\ 2015, \mnras, 446, 1283

%%Cuadra 2009: Massive black hole binary mergers within subparsec scale gas discs

\bibitem[Cuadra et al.(2009)]{Cuadra09} Cuadra, J., Armitage, P.~J., Alexander, R.~D., \& Begelman, M.~C.\ 2009, \mnras, 393, 1423

% Dermine 2013: Ecc pumping in post-agb stars

\bibitem[Dermine et al.(2013)]{Dermine13} Dermine, T., Izzard, R.~G., Jorissen, A., \& Van Winckel, H.\ 2013, \aap, 551, A50

%% D'Orazio et al 2013
\bibitem[D'Orazio et al.(2013)]{DOrazio13} D'Orazio, D.~J., Haiman, Z., \& MacFadyen, A.\ 2013, \mnras, 436, 2997

%% D'Orazio et al 2016
\bibitem[D'Orazio et al.(2016)]{DOrazio16} D'Orazio, D.~J., Haiman, Z., Duffell, P., MacFadyen, A., \& Farris, B.\ 2016, \mnras, 459, 2379

%% Dunhill and Alexander 2013

\bibitem[Dunhill \& Alexander(2013)]{Dunhill13} Dunhill, A.~C., \& Alexander, R.~D.\ 2013, \mnras, 435, 2328

%% Dunhill 2015: binary SMBH accretion

\bibitem[Dunhill et al.(2015)]{Dunhill15} Dunhill, A.~C., Cuadra, J., \& Dougados, C.\ 2015, \mnras, 448, 3545

%% Dvorak 1986: Stability
\bibitem[Dvorak(1986)]{Dvorak86} Dvorak, R.\ 1986, \aap, 167, 379

%% Escala et al 2005
\bibitem[Escala et al.(2005)]{Escala05} Escala, A., Larson, R.~B., Coppi, P.~S., \& Mardones, D.\ 2005, \apj, 630, 152

%% Farris et al 2014
\bibitem[Farris et al.(2014)]{Farris14} Farris, B.~D., Duffell, P., MacFadyen, A.~I., \& Haiman, Z.\ 2014, \apj, 783, 134

%% Goldreicg and Tremaine 1979: Excitation of density waves at resonances

\bibitem[Goldreich \& Tremaine(1979)]{GT79} Goldreich, P., \& Tremaine, S.\ 1979, \apj, 233, 857

%% Goldreich and Tremaine 1980: disc satellite interactions

\bibitem[Goldreich \& Tremaine(1980)]{GT80} Goldreich, P., \& Tremaine, S.\ 1980, \apj, 241, 425

%% Haisch 2001: Protoplanetary disc lifetimes

\bibitem[Haisch et al.(2001)]{Haisch01} Haisch, K.~E., Jr., Lada, E.~A., \& Lada, C.~J.\ 2001, \apjl, 553, L153

%% Holman + Wiegert 1999

\bibitem[Holman \& Wiegert(1999)]{HolmanWiegert99} Holman, M.~J., \& Wiegert, P.~A.\ 1999, \aj, 117, 621

%% Jetley 2008: ChaNGa paper

\bibitem[Jetley et al.(2008)]{Jetley08} Jetley, P., Gioachin, F., Mendes, C.,
Kale, L., Quinn, T.\ 2008, Proceedings of IEEE International Parallel and Distributed Processing Symposium 2008

%% Kale + Krishnan 1996
\bibitem[Kale \& Krishnan(1996)]{KaleKrishnan96} Kale, L. V. \& Krishnan, Sanjeev\ 1996, Parallel Programming using C++,
ed. Gregory V. Wilson and Paul Lu. (MIT Press), 175

%% Kley 2008: disc eccentricity

\bibitem[Kley et al.(2008)]{Kley08} Kley, W., Papaloizou, J.~C.~B., \& Ogilvie, G.~I.\ 2008, \aap, 487, 671

%% Kley 2014: Kepler 38 + planet migration

\bibitem[Kley \& Haghighipour(2014)]{Kley14} Kley, W., \& Haghighipour, N.\ 2014, \aap, 564, A72

%% Kley 2015

\bibitem[Kley \& Haghighipour(2015)]{KleyHag15} Kley, W., \& Haghighipour, N.\ 2015, \aap, 581, A20

%% Kostov 2015: New CB discovery paper

\bibitem[Kostov et al.(2015)]{Kostov15} Kostov, V.~B., Orosz, J.~A., Welsh, W.~F., et al.\ 2015, arXiv:1512.00189

%% Kozai 1962
\bibitem[Kozai(1962)]{Kozai62} Kozai, Y.\ 1962, \aj, 67, 591

%% Leung Lee 2013: Analytic Theory for orbits of circumbinary planets

\bibitem[Leung \& Lee(2013)]{LeungLee13} Leung, G.~C.~K., \& Lee, M.~H.\ 2013, \apj, 763, 107

%% Lidov 1962
\bibitem[Lidov(1962)]{Lidov62} Lidov, M.~L.\ 1962, \planss, 9, 719

%% Lines 2015: Modeling circumbinary protoplanetary discs

\bibitem[Lines et al.(2015)]{Lines15} Lines, S., Leinhardt, Z.~M., Baruteau, C., Paardekooper, S.-J., \& Carter, P.~J.\ 2015, \aap, 582, A5

%% Lines 2016: Modeling CBP discs II 

\bibitem[Lines et al.(2016)]{Lines16} Lines, S., Leinhardt, Z.~M., Baruteau, C., Paardekooper, S.-J., \& Carter, P.~J.\ 2016, arXiv:1604.01778

%% Lodato and Price 2010
\bibitem[Lodato \& Price(2010)]{Lodato10} Lodato, G., \& Price, D.~J.\ 2010, \mnras, 405, 1212

%% Lubow 1991: Superhump  eccentric discs

\bibitem[Lubow(1991)]{Lubow91} Lubow, S.~H.\ 1991, \apj, 381, 259 

%% Arty + Lubow 1996b: disc-binary interactions
\bibitem[Lubow \& Artymowicz(1996)]{Arty96b} Lubow, S.~H., \& Artymowicz, P.\ 1996, NATO Advanced Science Institutes (ASI) Series C, 477, 53

%% Lubow + Arty 2000: PPIV

\bibitem[Lubow \& Artymowicz(2000)]{Arty2000} Lubow, S.~H., \& Artymowicz, P.\ 2000, Protostars and Planets IV, 731

%% Lubow et al 2015
\bibitem[Lubow et al.(2015)]{Lubow15} Lubow, S.~H., Martin, R.~G., \& Nixon, C.\ 2015, \apj, 800, 96

%% MacFadyen 2008: An Eccentric Circumbinary Accretion disc and the Detection of Binary Massive Black Holes

\bibitem[MacFadyen \& Milosavljevi{\'c}(2008)]{MacFadyen08} MacFadyen, A.~I., \& Milosavljevi{\'c}, M.\ 2008, \apj, 672, 83

%% Martin et al 2014

\bibitem[Martin et al.(2014)]{Martin14} Martin, R.~G., Nixon, C., Lubow, S.~H., et al.\ 2014, \apjl, 792, L33

%% Mayer et al 2007

\bibitem[Mayer et al.(2007)]{Mayer07} Mayer, L., Kazantzidis, S., Madau, P., et al.\ 2007, Science, 316, 1874

%% Mazeh 2008

\bibitem[Mazeh(2008)]{Mazeh08} Mazeh, T.\ 2008, EAS Publications Series, 29, 1

%% Meibom 2005

\bibitem[Meibom \& Mathieu(2005)]{Meibom05} Meibom, S., \& Mathieu, R.~D.\ 2005, \apj, 620, 970

%% Menon 2015: ChaNGa paper

\bibitem[Menon et al.(2015)]{Menon15} Menon, H., Wesolowski, 
L., Zheng, G., et al.\ 2015, Computational Astrophysics and Cosmology, 2, 1

%% Meru and Bate 2012
\bibitem[Meru \& Bate(2012)]{Meru12} Meru, F., \& Bate, M.~R.\ 2012, \mnras, 427, 2022

%%Monaghan viscosity 1983

\bibitem[Monaghan \& Gingold(1983)]{Monaghan83} Monaghan, J.~J., \& Gingold, R.~A.\ 1983, Journal of Computational Physics, 52, 374

%% Murray 1996
\bibitem[Murray(1996)]{Murray96} Murray, J.~R.\ 1996, \mnras, 279, 402

%% Nelson and Papaloizou 2003

\bibitem[Nelson \& Papaloizou(2003)]{Nelson03} Nelson, R.~P., \& Papaloizou, J.~C.~B.\ 2003, \mnras, 339, 993

%% Nixon et al 2013
\bibitem[Nixon et al.(2013)]{Nixon13} Nixon, C., King, A., \& Price, D.\ 2013, \mnras, 434, 1946

%% Orosz 2012: Kepler 38 Discovery paper

\bibitem[Orosz et al.(2012)]{Orosz12} Orosz, J.~A., Welsh, 
W.~F., Carter, J.~A., et al.\ 2012, \apj, 758, 87

%% Paardekooper 2012: How not to build a tatooine... 10/10 title

\bibitem[Paardekooper et al.(2012)]{Paardekooper12} Paardekooper, S.-J., Leinhardt, Z.~M., Th{\'e}bault, P., \& Baruteau, C.\ 2012, \apjl, 754, L16

%% Papaloizou 2001

\bibitem[Papaloizou et al.(2001)]{Papaloizou01} Papaloizou, J.~C.~B., Nelson, R.~P., \& Masset, F.\ 2001, \aap, 366, 263

%% Pelupessy 2013: against in-situ formation

\bibitem[Pelupessy \& Portegies Zwart(2013)]{Pelupessy13} Pelupessy, F.~I., \& Portegies Zwart, S.\ 2013, \mnras, 429, 895

%% Pierens + Nelson 2007

\bibitem[Pierens \& Nelson(2007)]{PierensNelson07} Pierens, A., \& Nelson, R.~P.\ 2007, \aap, 472, 993

%% Pierens + Nelson 2013

\bibitem[Pierens \& Nelson(2013)]{PierensNelson13} Pierens, A., \& Nelson, R.~P.\ 2013, \aap, 556, A134

%% Popova et al 2013

\bibitem[Popova \& Shevchenko(2013)]{Popova13} Popova, E.~A., \& Shevchenko, I.~I.\ 2013, \apj, 769, 152

%% Pringle 1991

\bibitem[Pringle(1991)]{Pringle91} Pringle, J.~E.\ 1991, \mnras, 248, 754

%% Quinn et al 1997: multistepping
\bibitem[Quinn et al.(1997)]{Quinn97} Quinn, T., Katz, N., Stadel, J., \& Lake, G.\ 1997, arXiv:astro-ph/9710043

%% Raghavan 2010

\bibitem[Raghavan et al.(2010)]{Raghavan10} Raghavan, D., McAlister, H.~A., Henry, T.~J., et al.\ 2010, \apjs, 190, 1

%% Ragusa et al 2016

\bibitem[Ragusa et al.(2016)]{Ragusa16} Ragusa, E., Lodato, G., \& Price, D.~J.\ 2016, \mnras, 460, 1243

%% Roskar et al 2015
\bibitem[Ro{\v s}kar et al.(2015)]{Roskar15} Ro{\v s}kar, R., Fiacconi, D., Mayer, L., et al.\ 2015, \mnras, 449, 494

%% Richardson et al 2000

\bibitem[Richardson et al.(2000)]{Richardson2000} Richardson, D.~C., Quinn, T., Stadel, J., \& Lake, G.\ 2000, \icarus, 143, 45

%% Roedig 2011: Limiting binary eccentricity for binary SMBHs

\bibitem[Roedig et al.(2011)]{Roedig11} Roedig, C., Dotti, M., Sesana, A., Cuadra, J., \& Colpi, M.\ 2011, \mnras, 415, 3033

%% Roedig 2012

\bibitem[Roedig et al.(2012)]{Roedig12} Roedig, C., Sesana, A., Dotti, M., et al.\ 2012, \aap, 545, A127

%% Shi et al 2012
\bibitem[Shi et al.(2012)]{Shi12} Shi, J.-M., Krolik, J.~H., Lubow, S.~H., \& Hawley, J.~F.\ 2012, \apj, 749, 118

%% Silsbee 2015: Birth Locations of the Kepler Circumbinary Planets

\bibitem[Silsbee \& Rafikov(2015)]{Silsbee15a} Silsbee, K., \& Rafikov, R.~R.\ 2015, \apj, 808, 58

%% Toomre 1964: Q paper

\bibitem[Toomre(1964)]{Toomre64} Toomre, A.\ 1964, \apj, 139, 
1217

%% wadsley et al 2004: gasoline

\bibitem[Wadsley et al.(2004)]{Wadsley04} Wadsley, J.~W., Stadel, 
J., \& Quinn, T.\ 2004, \na, 9, 137

%% Welsh 2014: Kepler circumbinary planet results
\bibitem[Welsh et al.(2014)]{Welsh14} Welsh, W.~F., Orosz, J.~A., Carter, J.~A., \& Fabrycky, D.~C.\ 2014, IAU Symposium, 293, 125

%% Wisdom 1980: 3-body resonance overlap

\bibitem[Wisdom(1980)]{Wisdom80} Wisdom, J.\ 1980, \aj, 85, 1122 

%% Young and Clarke 2015

\bibitem[Young \& Clarke(2015)]{Young15} Young, M.~D., \& Clarke, C.~J.\ 2015, \mnras, 452, 3085

%% Zahn 1989: Tidal forces

\bibitem[Zahn \& Bouchet(1989)]{Zahn89} Zahn, J.-P., \& Bouchet, L.\ 1989, \aap, 223, 112
\end{thebibliography}
\end{document}